\documentclass[useAMS, usenatbib]{mnras}

\usepackage{graphicx}
\usepackage{natbib}
\usepackage{graphicx}
\usepackage{color}
\usepackage{pdfpages}
\usepackage{appendix}
\usepackage{subfigure}
\usepackage{color}
\usepackage{amsmath}
\usepackage{amssymb}

\begin{document}
%% define bibstyle and other definitions
\bibliographystyle{mnras}
%% renew commands
\renewcommand{\theenumi}{\arabic{enumi}.}

%% Codes
\def\araa{ARAA}
\def\nat{Nature}
\def\apjl{ApJ Letters}
\def\aapr{AAPR}
\def\actaa{ACTAA} 
\def\ssr{SSR}
\def\apj{ApJ}
\def\apjs{ApJs}
\def\apss{AP\&SS}
\def\pasp{PASP}
\def\aap{A\&A}
\def\aaps{A\&As}
\def\mnras{MNRAS}
\def\aj{AJ}
\def\rmxaa{RMXAA}
\def\memras{MmRA}
\def\baas{BAAS}
\def\nar{NAR}
\def\pasj{PASJ}
\def\pasa{PASA}
\def\memsai{MEMSAI}

%% define some control sequences for lines
\def\heiiopt{He~\textsc{ii}~$4686$~\AA}
\def\heiiuv{He~\textsc{ii}~$1640$~\AA}
\def\heiioptnew{He~\textsc{ii}~$3202$~\AA}
\def\civfull{C~\textsc{iv}~$1550$~\AA}
\def\oiiifull{[O~\textsc{iii}]~$5007$~\AA}
\def\oiiidoublet{[O~\textsc{iii}]~$4959$~\AA}
\def\fwh{FWHM[H$\beta$]}
\def\ewo{EW[O~\textsc{iii}]}
\def\la{Ly$\alpha$}
\def\ha{H$\alpha$}
\def\hb{H$\beta$}
\def\heiiline{He~\textsc{ii}~$4686$~\AA}

\def\ovi{O~\textsc{vi}}
\def\oiii{O~\textsc{iii}}
\def\civ{C~\textsc{iv}}
\def\nv{N~\textsc{v}}
\def\civ{C~\textsc{iv}}
\def\mgii{Mg~\textsc{ii}}
\def\mg{Mg~\textsc{ii}}
\def\hei{He~\textsc{i}}
\def\heii{He~\textsc{ii}}
\def\aliii{Al~\textsc{iii}}
\def\al{Al~\textsc{iii}}
\def\ciii{C~\textsc{iii}}
\def\ellp{{\ell^\prime}}
\def\up{{u^\prime}}
\def\cc{{\cal C}}
\def\hh{{\cal H}}
\def\ept{$\epsilon(\theta)$}
\def\eptl{$\epsilon_{line}(\theta)$}
\def\al{Al~\textsc{iii}}
\def\civline{C~\textsc{iv}~$1550$~\AA}
\def\nvline{N~\textsc{v}~$1240$~\AA}
\def\mgline{Mg~\textsc{ii}~$2800$~\AA}
\def\ciiiline{C~\textsc{iii}]~$1909$~\AA}

\def\top{\textsc{Topbase}}
\def\py{\textsc{Python}}
\def\tar{\textsc{Tardis}}
\def\cld{\textsc{Cloudy}}
\def\agn{\textsc{Agnspec}}
\def\kerrtrans{\textsc{Kerrtrans}}

\def\ew{{\rm EW}}

\newcommand{\EXPN}[2]{\mbox{$#1\times 10^{#2}$}}
\newcommand{\EXPU}[3]{\mbox{\rm $#1 \times 10^{#2} \rm\:#3$}}  % exponent with units
\newcommand{\POW}[2]{\mbox{$\rm10^{#1}\rm\:#2$}}
\def\LUM{\:{\rm erg\:s^{-1}}}
\def\FLUX{\:{\rm erg\:cm^{-2}\:s^{-1}}}
\def\OIGS{\:{\rm erg\:cm^{-2}\:s^{-1}\:\AA^{-1}}}

\def\Msol{$M_{\odot}\ $}
\def\mdote{\.m$\mathrm{_{Edd}}$}
\def\mdotes{\.m$\mathrm{_{Edd}~}$}
\def\mcrit{\.m$\mathrm{_{crit}}$}
\def\rg{$R_G$}
\def\mdot{\dot{m}}
\def\ept{$\epsilon(\theta)$}
\def\fbal{$f_{\mathrm{BAL}}$}

\title
[Quasar emission lines as probes of orientation]
{Quasar emission lines as probes of orientation:
implications for disc wind geometries and unification}

\author[Matthews et al.]{J.~H.~Matthews$^{1}$\thanks{james.matthews@physics.ox.ac.uk}, C.~Knigge$^2$ and K.~S.~Long$^{3,4}$
\medskip  
\\$^1$Astrophysics, Department of Physics, University of Oxford, Keble Road, Oxford, OX1 3RH, UK
\\$^2$School of Physics and Astronomy, University of Southampton, Highfield, Southampton, SO17 1BJ, UK
\\$^3$Space Telescope Science Institute, 3700 San Martin Drive, Baltimore, MD, 21218
\\$^4$Eureka Scientific, Inc., 2452 Delmer Street Suite 100, Oakland, CA 94602-3017
}

\date{Accepted 2017 January 24. Received 2017 January 16; in original form 2016 November 11.}

%%%%%%%%%%%%%%%%%%%%%%%%%%%%%%%%%%%%%%
%
%          ABSTRACT
%
%%%%%%%%%%%%%%%%%%%%%%%%%%%%%%%%%%%%%%%
\maketitle

\begin{abstract} 
The incidence of broad absorption lines (BALs) in quasar samples is 
often interpreted in the context of a geometric unification model consisting
of an accretion disc and an associated outflow.
We use the the Sloan Digital Sky Survey (SDSS) quasar sample to 
test this model by examining the equivalent widths (EWs) of 
\civline, \mgline, \oiiifull\ and \ciiiline. 
We find that the emission line EW distributions in BAL and non-BAL
quasars are remarkably similar -- a property that is inconsistent with scenarios in which a BAL
outflow rises equatorially from a geometrically thin, optically thick accretion disc.
We construct simple models to predict the distributions from various 
geometries; these models confirm the above finding and disfavour equatorial geometries. 
We show that obscuration, line anisotropy and general relativistic effects on 
the disc continuum are unlikely to hide an EW inclination dependence. We carefully examine the radio and polarisation properties of 
BAL quasars. Both suggest that they are most likely viewed (on average) from intermediate inclinations, between type 1 and type 2 AGN. 
We also find that the low-ionization BAL quasars in our sample are not confined to one region of 
`Eigenvector I' parameter space. Overall, our work leads to one of the following conclusions, or 
some combination thereof: (i) the continuum does not emit like a geometrically thin, optically thick disc;
(ii) BAL quasars are viewed from similar angles to non-BAL quasars, i.e. low inclinations; 
(iii) geometric unification does not explain the fraction of BALs in quasar samples.
\end{abstract}

\section{Introduction}

The ultraviolet (UV) and optical spectra of type 1 quasars are characterised
by a blue continuum and a series of broad and narrow emission lines. Approximately 20\% of quasars also show blue-shifted, 
broad absorption lines (BALs) in their UV spectra \citep{weymann1991,knigge2008,allen2011},
providing clear evidence that outflowing material intersects the line of sight to
the continuum source. Most BAL quasars (BALQs) 
exhibit only high ionization BALs (HiBALs), but a
subset ($\sim10\%$) also show absorption in lower ionization species such as Mg~\textsc{ii} and are known as LoBAL quasars
\citep[LOBALQs; e.g.][]{reichard2003}.

The BAL phenomenon is normally explained either by evolutionary 
models \citep{gregg2000,becker2000,gregg2006,farrah2007,lipari2009}, 
in which quasars spend $\sim20\%$ of their lifetime as 
BALQs, or by a geometric interpretation, in which the
BAL fraction roughly corresponds to the covering factor of an ever-present wind 
\citep{MCGV95, elvis2000}.
In the latter case, winds could also create the broad emission lines 
seen in quasar spectra, meaning they offer a natural avenue through which the diverse
phenomenology of quasars can be {\em unified} according to orientation. 
This principle of geometric unification
is not confined to disc wind models; orientation-based models have also been famously
invoked to explain the type 1/type 2 and radio-loud/radio-quiet dichotomies
in AGN \citep{antonucci1985,UP95}, as well as the `Eigenvector I' trend in quasars
\citep{borosongreen,marziani2001,shenho2014}.

\newpage
Geometric unification scenarios require -- by definition -- that different classes
of objects are viewed from different angles. They thus predict that any orientation-dependent 
observable should vary accordingly between the classes. Empirical (albeit model-dependent)
examples include the differences in polarisation properties \citep[e.g.][]{marin2014}
and absorbing column densities \citep[e.g.][]{lusso2012} between type 1 and type 2 AGN.
However, in general, obtaining
reliable orientation indicators in quasars and AGN is difficult 
\citep[see ][for a summary]{marin2016}. 
Perhaps as a result of this problem, 
directly opposing geometries have been proposed for 
BAL outflows themselves. Polarisation studies imply that the 
wind is roughly equatorial \citep{goodrich1995, cohen1995,lamy2004,brotherton2006}, 
as also suggested by hydrodynamical and radiative transfer simulations 
\citep{PSK2000,PK04, higginbottom2013, borguet2010}.
However, there is also evidence for polar BAL outflows in 
radio-loud (RL) sources \citep{zhou2006,ghoshpunsly2007}.

One potential orientation indicator is the equivalent width
(EW) of the emission lines 
\citep[e.g.][]{risaliti2011}. 
The UV-optical continuum in AGN, 
known as the big blue bump, is normally
thought to originate from a geometrically 
thin, optically thick accretion disc 
surrounding the central black hole
\citep[e.g.][]{shields1978,malkan1982,malkan1983,capellupo2015}.
The emission from this disc should be 
strongly anisotropic due
to foreshortening and possibly 
limb darkening \citep[e.g.][see also section~\ref{sec:disc_agn}]{herter1979,wade1984,laor1989,hubeny2000}.
On the other hand, there is no 
{\em a priori} reason why emission lines
should emit in the same manner as the 
disc. If a line is at all optically thin, 
or formed in a region with isotropic
escape probabilities, then the 
emission will radiate isotropically
(see section~\ref{sec:line_aniso}).
It follows that we might expect 
emission line EW to increase with 
inclination and be largest for edge-on 
systems, especially for optically 
thin narrow emission lines.

The variation of EW with inclination is demonstrated neatly by the behaviour 
of emission lines in high-state accreting white dwarfs (AWDs), often thought to be 
reasonable quasar analogues. In these systems, 
inclinations are more well constrained and a geometrically thin, optically thick
accretion disc is established as the continuum source 
\citep[][and references therein]{warner1995}.
High-state AWDs do indeed show a clear trend of increasing line EW with inclination 
\citep{hessman1984,patterson1984,echevarria1988}. 
This behaviour is also seen in radiative transfer simulations in both AWDs and 
quasars \citep{noebauer,M15,M16}. 
The upshot is that quasar line EWs could potentially be used 
(i) to test geometric unification models for, e.g., quasars and the BAL phenomenon, 
or (ii) to help {\em understand the origin of the UV-optical continuum in AGN}. The latter
is particularly important given the currently unsatisfactory understanding of 
the quasar continuum source \citep[e.g.][see also section~\ref{sec:discuss}]{koratkar1999}.

\begin{figure} %fullpage
\centering
\includegraphics[width=0.5\textwidth]{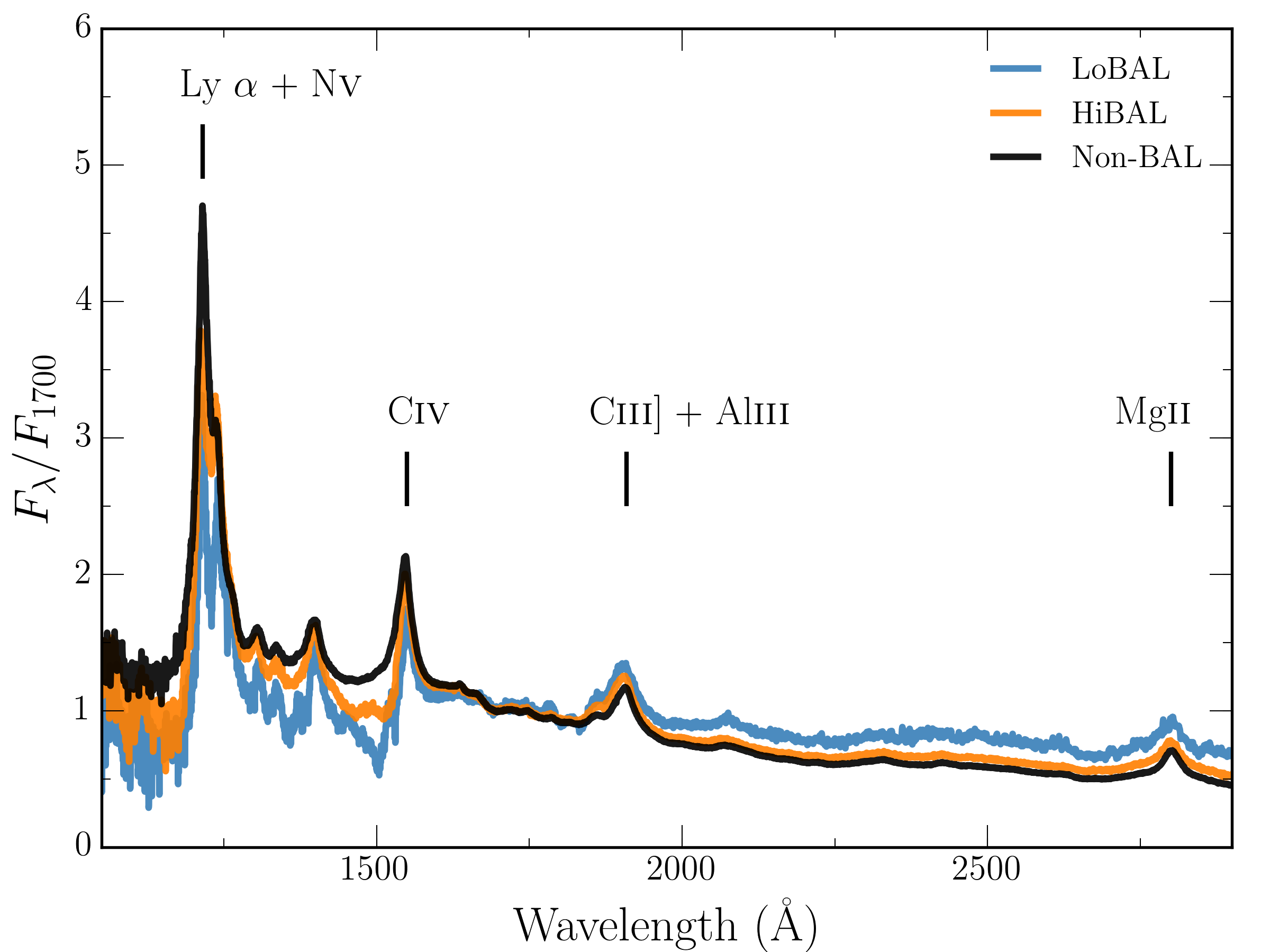}
\caption
{
Sloan Digital Sky Survey composite spectra for 
HiBAL, LoBAL and non-BAL quasars from Reichard et al. (2003).
The prominent emission lines are marked and the spectra are
normalised to the monochromatic flux at $1700$\AA.
}
\label{fig:composites}
\end{figure} %fullpage

\begin{table*}
\centering
\caption
{
A summary of the data samples described in the text. HiBAL
identification is not possible in sample A due to the wavelength coverage.
}
\begin{tabular}{p{1cm} p{1cm} p{2cm} p{2.2cm} p{6cm} p{1.2cm}}
\hline
Sample & Size & $N_{\mathrm{BAL}}$ & Redshift range & Lines Used & Source \\
\hline \hline 
A & 16,742 & 58 (all LoBAL) & $0.35<z<0.83$ & \mgline, \oiiifull\ & S11/DR7 \\
B & 80,429 & 6744 & $1.45<z<2.28$ & \mgline, \ciiiline, \civline\ & P16/DR12 \\
\hline 
\end{tabular}
\label{tab:samples}
\end{table*}

The ideal emission line to use for this method would be one that is completely isotropic,
i.e. {\em optically thin}. The \oiiifull\ narrow emission line fulfils this criteria, since it
is a strong, forbidden line formed in the narrow-line region (NLR) of AGN.
Any dispersion in the distribution of \oiiifull\ EW (\ewo) must therefore be driven by some 
combination of the intrinsic luminosity \citep{borosongreen}, 
the covering factor/geometry of the NLR \citep{baskin2005} and the 
inclination of the disc \citep{risaliti2011}. 
In a recent study, \citet[][hereafter R11]{risaliti2011} showed that 
the \ewo\ distribution had a high power law tail and
could be well fitted by a simple model driven purely 
by disc inclination.
In this study, we compare the EWs of a number of emission lines in BAL and non-BAL quasar samples
in order to test unification models in which the continuum source 
is a geometrically thin, optically thick accretion disc. Our study is motivated by the remarkably 
similar emission line properties of BAL and non-BAL quasars (see Fig.~\ref{fig:composites})
-- a similarity that would not be expected from simple models in which BALQs are viewed 
from equatorial angles.

This paper is structured as follows. First, we describe
the data sample and selection criteria being used. We begin by
simply examining the BAL and non-BAL quasar distributions for 
four emission lines: the narrow \oiiifull\ line, and the broad
\civfull, \ciiiline\ and \mgline\ lines. We construct some simple
toy geometric models in which an optically thick, geometrically thin
disc acts as the continuum source, and BAL quasars are viewed from some subset
of angles dependent on the geometry of the BAL outflow. We find, as expected,
that such toy models predict large differences in EW distributions if 
BALs are only seen at high inclinations. Our discussion begins
in section~\ref{sec:discuss}, in which we first explore if there are any 
straightforward explanations -- namely general relativistic effects, 
line anisotropy or obscuration -- that could readily explain the EW distribution
and allow BAL quasars to be seen at systematically different inclinations
to non-BAL quasars. In section 4, we discuss the results in the context of radio and 
polarisation measurements of AGN, and explore the location of BAL quasars 
in `Eigenvector 1' parameter space. Finally, in section~\ref{sec:ew_conclusions}, 
we summarise our results.

%%%%%%%%%%%%%%%%%%%%%%%%%%%%%%%
% RESULTS
%%%%%%%%%%%%%%%%%%%%%%%%%%%%%%%

\section{The EW Distributions of BAL and non-BAL Quasars}
\label{sec:results}

Our data samples are based on two catalogs from
the Sloan Digital Sky Survey (SDSS):
The \citet[][hereafter S11]{shen2011} catalog of
105,783 quasars from the SDSS
Data Release (DR) 7 \citep{sdssdr7} and the \citet[][hereafter P16]{paris2016}
catalog of 297,301 quasars from the SDSS DR 12. 
As we will use emission line diagnostics in this study,
our samples must be divided according to which 
emission lines are present in 
the SDSS wavelength range at a given redshift. 
Sample A contains all quasars from S11\footnote{Note that although \cite{albareti2015} present a more
recent compendium of \oiiifull\ line measurements based on DR12, the 
catalog actually contains less quasars in total than S11, so we revert 
to DR7 for sample A. This is due to the coverage of the samples at $z<0.83$ 
\citep{paris2016}.} 
within the redshift range $0.35<z<0.83$, 
such that the \mgline\ and \oiiifull\ line EWs are both measured, 
and \mg\ LoBAL identification
is possible.  Sample B contains all quasars from P16 
within the redshift range $1.45<z<2.28$, such that 
the EWs and presence of BALs in \mgline, \ciiiline\ 
and \civline\ are both measurable. The samples are summarised 
in Table~\ref{tab:samples}.

S11 are careful to take into account traditional problems with quasar line fitting,
such as narrow line or Fe pseudocontinuum contamination, in their fits to 
emission line profiles and resultant EW measurements. For \mgline\
this includes careful subtraction of the nearby Fe emission using the \cite{vestergaard2001}
templates. This subtraction is not included for \civfull,
as the Fe emission is less prominent and harder to model. This may lead to
a systematic overestimate by $\sim0.05$ dex in the \civ\ line EW. 
The \oiiifull\ line is fitted with a Gaussian. The flux ratio of this line 
with the sister component of the doublet, \oiiidoublet, is found to agree well 
with the theoretical expectation of around 3, implying a reliable subtraction 
of broad \hb. In order to mask out the effects of e.g., absorption, 
on the \civ\ and \mg\ lines, 
S11 ignore $3\sigma$ outliers in the fit to the profile. 
Although P16 provide less detail on the emission 
line fitting process, we have verified that the EWs of emission lines measured
in both P16 and S11 have very similar means and variances, although some 
differences in shape occur, possibly due to the changes in quasar target 
selection (see P16). Adopting a different SDSS quasar catalog does therefore
not affect any of our conclusions. 
The quasar selection criteria for the SDSS DR7 and DR12 
are described by \cite{schneider2010} and \cite{ross2012} respectively. 
In S11, the quasars must have at least one broad line whilst in our 
P16 sample we only consider broad emission lines;
the non-BAL quasars discussed in the next section 
are thus unobscured, type 1 quasars with broad emission lines. 
This means we do not consider obscured, type 2 quasars in our analysis, 
although we note that the
the geometry of the model discussed in section~\ref{sec:toymods} 
could contribute to type 2 observational biases
\citep{reyes2008,alexandroff2013,yuan2016}.

Based on all the above considerations, 
the S11 and P16 catalogs make for a reliable set of EW measurements.
This is especially true when making inferences from 
multiple emission lines, as systematics inherent to individual lines 
or spectral windows are less likely to affect the analysis as a whole.
Importantly, the data samples chosen allow for emission lines formed in 
different regions (the broad-line region [BLR] and NLR) and by different atomic 
transitions (forbidden, intercombination and permitted dipole) 
to be studied. 

\subsection{The Observed EW Distributions}

\begin{table*}
\centering
\caption
{
The mean ($\mu_\mathrm{EW}$), median ($m_\mathrm{EW}$)
and standard deviation ($\sigma_\mathrm{EW}$) of each EW distribution
used in this study, for both BAL and non-BAL quasars. The distributions 
are shown in Fig.~\ref{fig:ew_hists}. The quantity $\sigma_m$ gives the
(asymmetric) distances either side of the median to the 16th and 84th percentiles of the 
cumulative distribution function, thus enclosing $68\%$ of the total counts.
Units are in \AA.
%except for $p_{\mathrm{KS}}$, which is a statistical $p$-value
All values are given to two-decimal places. 
}
\label{mean_table}
\begin{tabular}
{p{0.7cm} p{2cm} p{1cm} p{1cm} p{1cm} p{1cm} p{0.2cm} p{1cm} p{1cm} p{1cm} p{1cm} p{0.1cm} p{2cm}}
\hline

\multicolumn{2}{}{} & 
\multicolumn{4}{c}{non-BAL} & & \multicolumn{4}{c}{BAL} & & \\
Sample & Line &
$\mu_\mathrm{EW}$ & $m_\mathrm{EW}$ & $\sigma_\mathrm{m}$ & $\sigma_\mathrm{EW}$ & &
$\mu_\mathrm{EW}$ & $m_\mathrm{EW}$ & $\sigma_\mathrm{m}$ & $\sigma_\mathrm{EW}$ & &
$\Delta \mu_\mathrm{EW}$ \\
\hline \hline 
A & [O~\textsc{iii}]~$5007$\AA & 25.20 & 15.70 & $_{-8.78}^{+21.45}$ & 36.92 & & 26.76 & 12.30 & $_{-9.28 }^{+32.37}$ & 39.25 & & $1.56\pm5.16$ \\
A & Mg~\textsc{ii}~$2800$\AA & 45.74 & 36.80 & $_{-13.36}^{+24.86}$ & 41.42 & & 32.55 & 27.80 & $_{-10.42 }^{+16.87}$ & 16.90 & & $-13.19\pm2.26$ \\
B & C~\textsc{iii}]~$1909$\AA & 38.13 & 33.62 & $_{-10.71}^{+16.50}$ & 22.45 & & 34.67 & 32.57 & $_{-9.13 }^{+12.09}$ & 14.67 & & $-3.46\pm0.21$ \\
B & Mg~\textsc{ii}~$2800$\AA & 52.40 & 45.01 & $_{-16.82}^{+25.79}$ & 40.35 & & 41.39 & 37.15 & $_{-12.61 }^{+17.80}$ & 21.64 & & $-11.01\pm0.31$ \\
B & C~\textsc{iv}~$1550$\AA & 48.51 & 40.86 & $_{-15.74}^{+28.32}$ & 29.81 & & 36.68 & 34.74 & $_{-11.30 }^{+13.12}$ & 15.09 & & $-11.83\pm0.25$ \\
\hline 
\end{tabular}
\end{table*}

\begin{figure*} %fullpage
\centering
\includegraphics[width=0.95\textwidth]{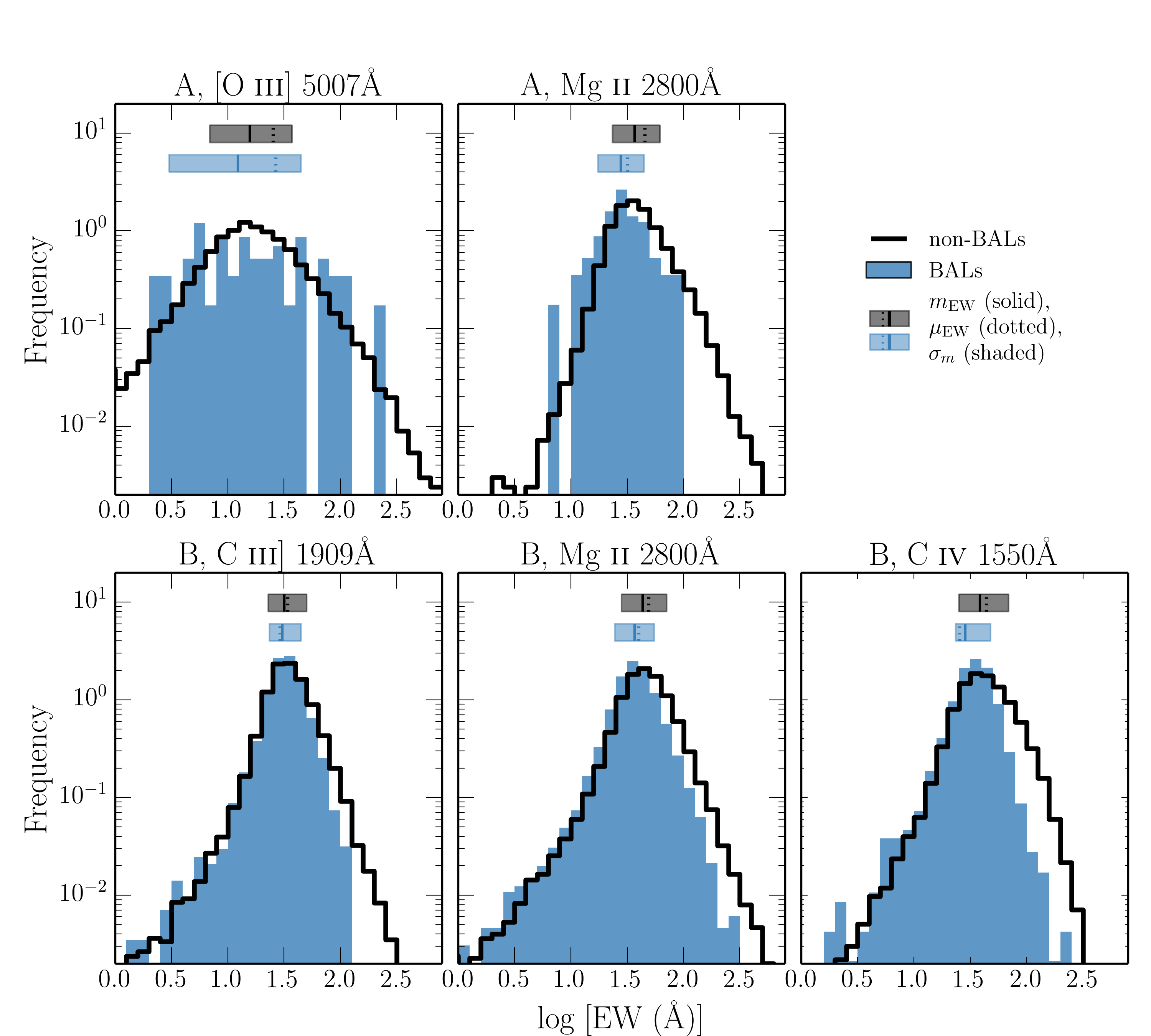}
\caption
{
Normalised histograms of equivalent width for different emission 
lines from the two different samples. The top two panels show 
the distributions from sample A for the forbidden \oiiifull\ 
line and permitted dipole transition \mgline. The bottom three 
panels show the distributions from sample B for 
the semi-forbidden/intercombination line \ciiiline\ and the permitted 
dipole transitions \mgline\ and \civline. In all cases, the non-BAL 
quasar distributions are plotted with a black line and the BALQs 
with a solid blue histogram. The binning is logarithmic. 
The mean ($\mu_\mathrm{EW}$), median ($m_\mathrm{EW}$)
and $\sigma_m$ (defined in the text).
}
\label{fig:ew_hists}
\end{figure*} %fullpage

It is apparent from the composite spectra of BAL and non-BAL quasars (Fig.~1)
that, when one compensates for how the blue-shifted absorption affects 
the composites, BAL and non-BAL quasars seem to possess very 
similar emission line properties. 
This has been noted by, e.g., \cite{weymann1991} in the past. 
Composite spectra could, however, hide differences between the two 
populations since they are built from a geometric mean \citep{reichard2003}. 
We thus show histograms of EWs for a number of different emission lines in 
Fig.~\ref{fig:ew_hists}. We give the mean, median and 
standard deviations of each of these distributions  
in Table~\ref{mean_table}, as well as the quantity $\sigma_m$, which gives the
(asymmetric) distances either side of the median to the 16th and 84th 
percentiles of the cumulative distribution function.
This quantity therefore encloses $68\%$ of the total counts.  We also show
the difference between the means of the BAL and non-BAL distributions.
We mark $\sigma_m$, the mean and the median on Fig.~\ref{fig:ew_hists}. 
In addition, we performed a two-tailed Kolgomorov-Smirnov (KS) 
test on each of the samples. 
The $p$-values from these tests are very small in each case. The largest
$p$-value obtained is $0.0027$ for the smallest sample ([\oiii]), 
showing that we can reject the KS null hypothesis at $>3\sigma$ in each case.
However, this gives no real information about orientation as the KS null 
hypothesis will also be rejected when they are small differences in the 
intrinsic populations and even when the BALQ EWs are systematically 
{\em lower} than non-BAL quasar EWs.This means that the KS test has 
very limited use in this circumstance.

The EW of an isotropic line is related to the intrinsic, 
`face-on' equivalent width, $\ew_*$ by the equation
\begin{equation}
\ew = \ew_* /\epsilon(\theta)
\end{equation}
where $\theta$ is the viewing angle with respect to the symmetry axis 
and $\epsilon(\theta)$ is the `angular emissivity function', which describes 
how the continuum luminosity from the disc varies as a function of viewing angle.
For a foreshortened disc this is simply $\epsilon(\theta) = \cos \theta$. 
Note that isotropic line emission may not be a reasonable assumption for optically thick
permitted dipole transitions and possibly even semi-forbidden intercombination lines
such as \ciiiline\ \citep{bhatia1992}; the effect of line anisotropy is discussed 
further in section~\ref{sec:line_aniso}.

If BALQs are preferentially viewed from larger-than-average angles,
we would expect them to possess higher EWs. 
As shown in Fig.~\ref{fig:ew_hists} and Table~\ref{mean_table}, 
the BALQ mean EW values are not higher than 
for non-BAL quasars -- in fact, in most cases they are lower. 
Furthermore, the distribution
shapes are generally very similar. 
Similar EW distributions are  
not expected from a model in which the continuum comes from a foreshortened disc, 
and BAL outflows are at all equatorial. To examine this apparent discrepancy
more concretely, 
we now devote some time to simulating the expected BAL and 
non-BAL EW distributions from different unification geometries.

\subsection{Expected EW Distributions: Toy Models}
\label{sec:toymods}

\begin{figure*}
\centering
\includegraphics[width=0.5\textwidth]{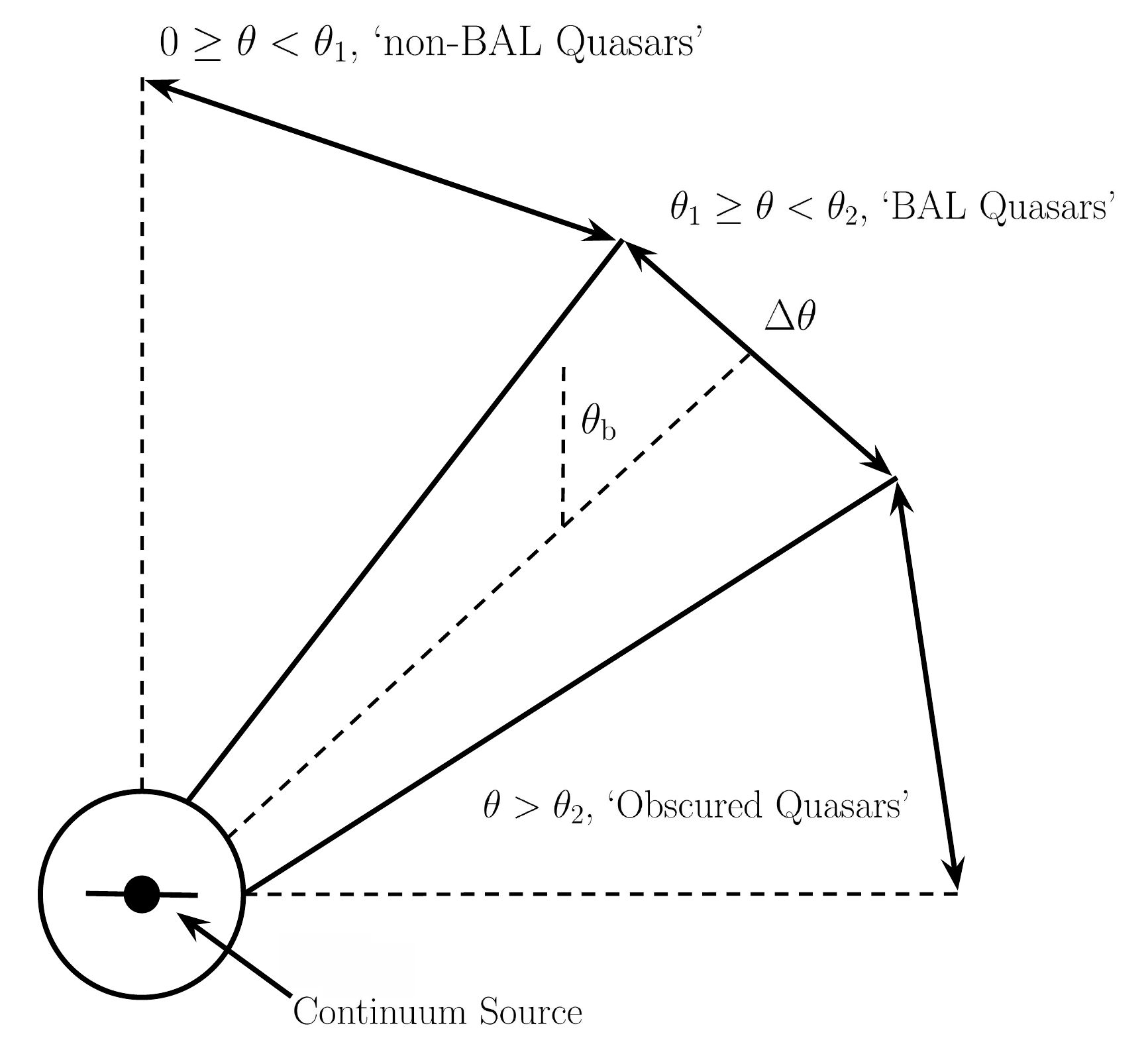}
\caption
{
The geometry of the toy model used to carry out the numerical simulations.
The marked angles and designations are described in the text.
}
\label{fig:cartoon}
\end{figure*}

\begin{figure*}
\centering
\includegraphics[width=1.0\textwidth]{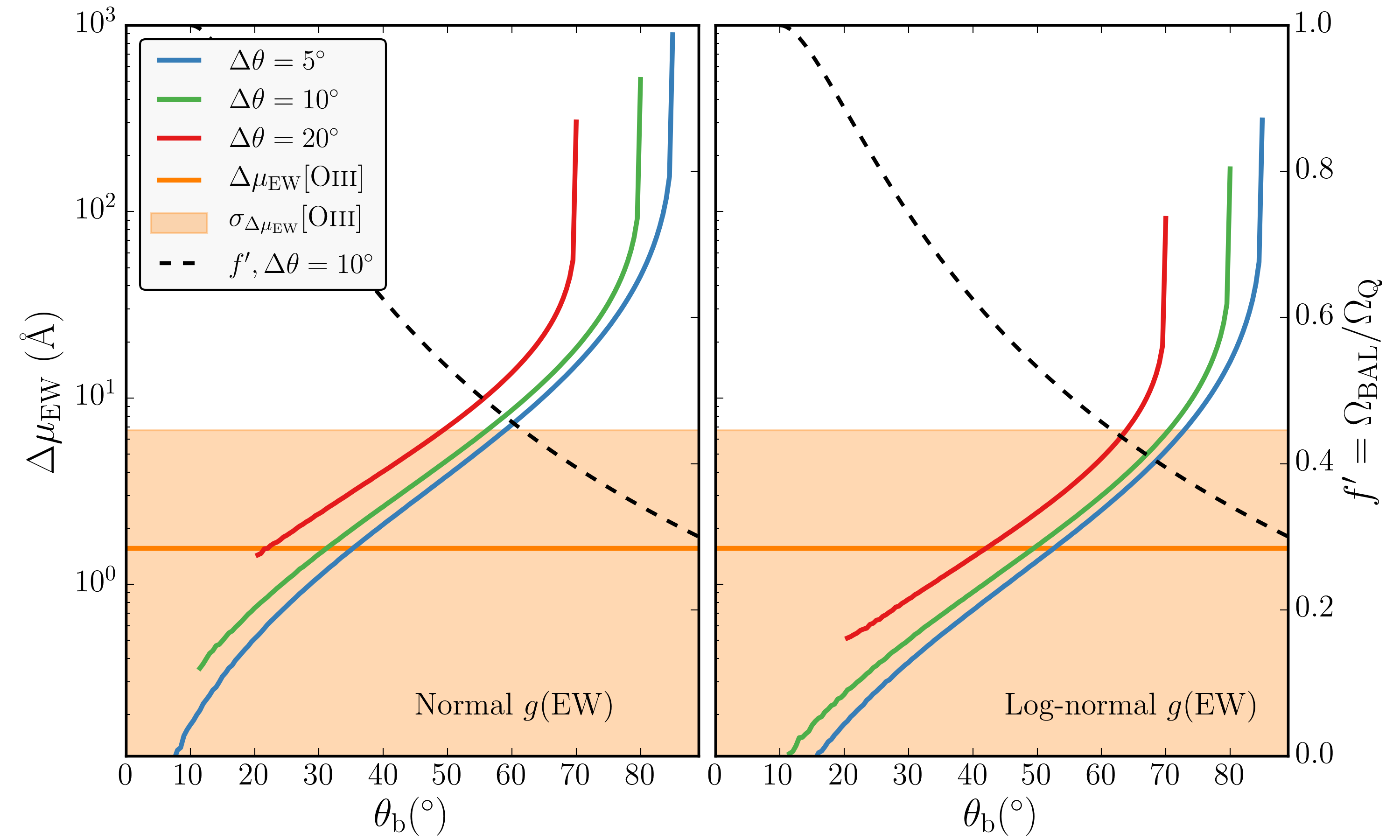}
\caption
{
The results of the numerical simulation described in section~\ref{sec:toymods}, 
showing difference in the mean EW of the non-BAL and BAL mock samples, $\Delta \mu_{\mathrm{EW}}$,
as a function of the bending angle, $\theta_{\mathrm{b}}$.
Results are shown for three different outflow covering factors ($\Delta \theta = 5^\circ,10^\circ,20^\circ$) and for 
Normal (left) and Log-normal (right) forms of $g(\mathrm{EW})$.
The Normal distribution has $\mu_*=10$, $\sigma_*=5$ and the 
Log-normal distribution has $\mu_*=1$, $\sigma_*=0.7$.
The orange 
horizontal line shows the {\em observed} value of 
$\Delta \mu_{\mathrm{EW}}=1.56$ for the 
\oiiifull\ emission line, while the shaded region shows the $1\sigma$ error
($\pm 5.16$) on that value -- note that the error extends to negative values 
which cannot be shown on these axes. 
The dashed black line, plotted on a different axis, shows the ratio of the 
outflow covering factor 
($\theta_1 \rightarrow \theta_2$) to the total quasar covering factor 
($0 \rightarrow \theta_2$) for $\Delta \theta = 10^\circ$.
}
\label{fig:deltamu}
\end{figure*}

A schematic showing the geometry used in our numerical simulations is
shown in Fig.~\ref{fig:cartoon}. In order to test geometric unification
scenarios such as that proposed by \cite{elvis2000}, we assume that BALs 
can be seen in spectra when the viewing angle with respect to the
symmetry axis is between $\theta_1$ and $\theta_2$. Non-BAL quasars
are seen when $\theta < \theta_1$. Beyond $\theta_2$
we assume the object is obscured and does not appear in the quasar
sample, as it only consists of type 1 objects. 
We also define the `bending angle' of the BAL outflow, 
which is simply 
\begin{equation}
\theta_{\mathrm{b}} = (\theta_1 + \theta_2) /2,
\end{equation}
and the opening angle, which is $\Delta \theta = (\theta_2-\theta_1)$.
These angles are marked in Fig.~\ref{fig:cartoon} and are similar to those
used by \cite{elvis2000} and \cite{marin2013} to describe equivalent geometries.

We carried out the following procedure to simulate
the effect of inclination on the EW distributions.
This method is similar to that used by R11
to fit the observed \ewo\ distribution in quasars. However, this is not an
attempt to fit the data, merely to demonstrate the expected geometric trends.
\begin{enumerate}
	\setlength\itemsep{1em}
	\item An isotropic angle was chosen for the mock quasar. 
	If $\theta<\theta_{1}$ then the mock quasar was designated as a non-BAL quasar.
	If $\theta_1 < \theta \leq \theta_2$, the mock quasar was designated as a BAL quasar
	and otherwise the object was ignored.
	\item For each mock sample, an EW$_*$ was drawn from an intrinsic 
	(i.e. `face-on') EW distribution for quasars, $g(\mathrm{EW})$. 
    To test both symmetric and asymmetric intrinsic distributions,
    this was assumed to take the form of a Normal or Log-normal distribution, following
    R11. \footnote{The mean, $\mu_*$, and width, $\sigma_*$, of the $g(\mathrm{EW})$ distribution
	is set by hand to give a reasonable approximation to the observed non-BAL quasar distribution, but we have also verified that the exact shape does not have a significant 
	effect on our conclusions. We adopt
    $\mu_*=10$, $\sigma_*=5$ for the Normal distribution, and $\mu_*=1$, $\sigma_*=0.7$
    for the Log-normal distribution.}
	\item The EW for each mock quasar 
	was estimated such that $\ew = \ew_* / \epsilon(\theta)$,
	and this process was repeated to build up a mock sample of $10^7$ objects.
	\item The difference between the mean EW of the non-BAL and BAL mock samples, 
	$\Delta \mu_{\mathrm{EW}}$, was recorded for each geometry.
\end{enumerate}
The results of this experiment are presented in Fig.~\ref{fig:deltamu}, where
we show $\Delta \mu_{\mathrm{EW}}$ as a function of $\theta_{\mathrm{b}}$ 
for three different values of $\Delta \theta$. Results for both Normal and 
Log-normal forms of $g(\mathrm{EW})$ are shown. We also show, with a horizontal
line, the {\em observed} value of $\Delta \mu_{\mathrm{EW}}$ for the 
\oiiifull\ emission line, while the shaded region shows the $1\sigma$ error
on this value. This is the most appropriate emission line to consider, as
it is forbidden and locally isotropic, but is also the most conservative, as
all the other emission lines considered have  
negative values of $\Delta \mu_{\mathrm{EW}}$ with a smaller error. 
On the same plot 
we also show $f^\prime$, which is given by
\begin{equation}
f^\prime = \frac{\Omega_{\mathrm{BAL}}}
{\Omega_{\mathrm{Q}}}
= \frac{\int^{\theta_2}_{\theta_1} d \Omega}
{\int^{\theta_2}_{0} d \Omega}
\end{equation}
where $\Omega_{\mathrm{BAL}}$ is the BAL quasar covering factor and 
$\Omega_{\mathrm{Q}}$ is the total quasar covering factor. 
This illustrates how the intrinsic BAL fraction depends on the outflow geometry, although it ignores flux selection effects (see section~\ref{sec:flux_select}).

\begin{figure*} %fullpage
\centering
\includegraphics[width=1.0\textwidth]{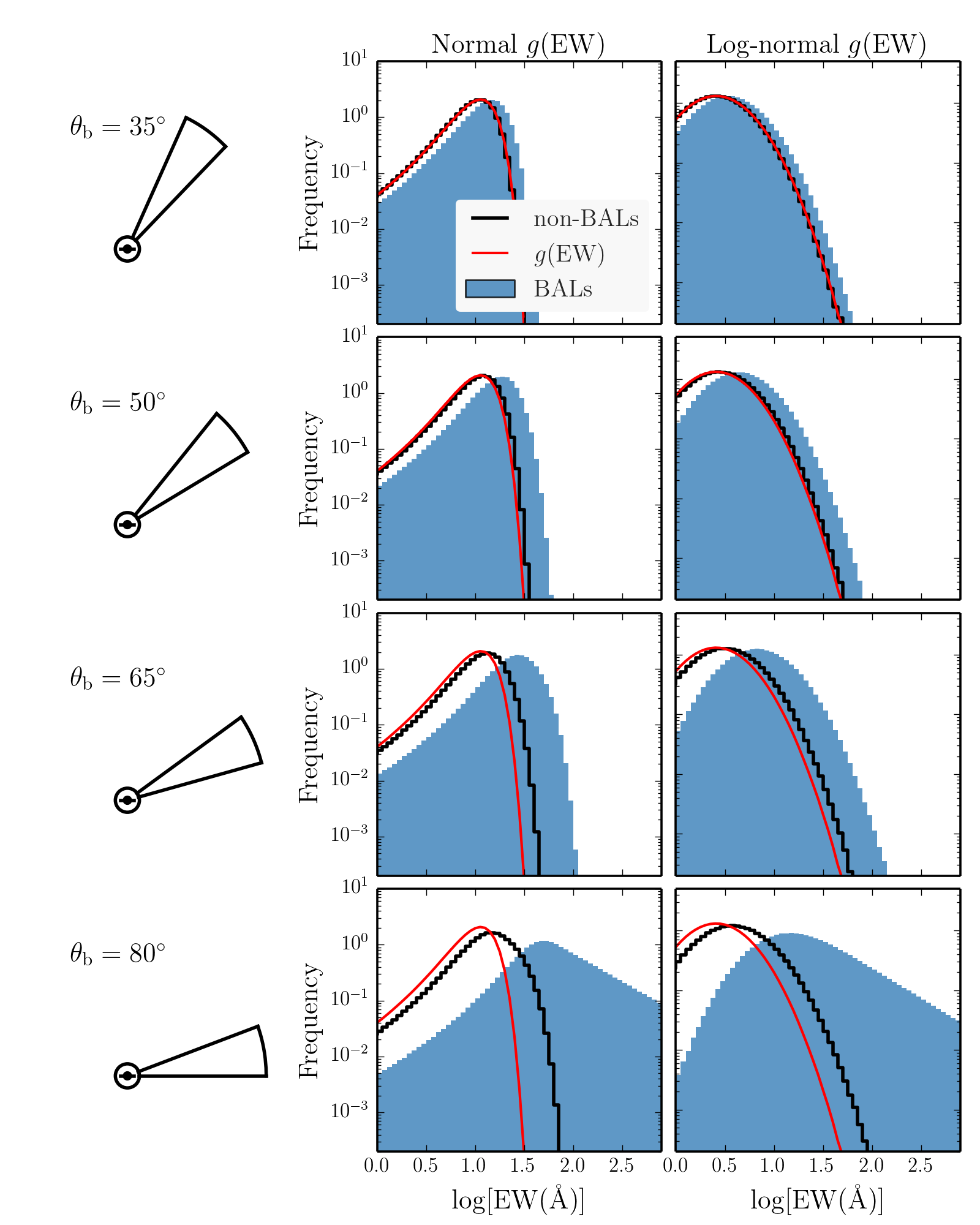}
\caption
{
Normalised histograms of mock equivalent widths for 
BAL and non-BAL quasars from the toy model
described in section~\ref{sec:toymods}, 
for four different values of 
$\theta_{\mathrm{b}}$ and 
$\Delta \theta = 20^\circ$. 
The intrinsic `face-on' distribution
used as input, $g(\mathrm{EW})$, 
is shown in red in each case and the BAL 
and non-BAL mock quasar 
samples are shown in the same corresponding 
colours as for the real data in 
Fig.~\ref{fig:ew_hists}.
The Normal $g(\mathrm{EW})$ distribution has 
$\mu_*=10$, $\sigma_*=5$ and the 
Log-normal distribution has $\mu_*=1$, 
$\sigma_*=0.7$.
}
\label{fig:ew_with_cartoon}
\end{figure*}

The toy model predicts that large differences in EW should be present 
between BAL and non-BAL samples for equatorial outflows, with $\Delta \mu_{\mathrm{EW}}>10$\AA\
expected for $\theta_{\mathrm{b}} \gtrsim 60^\circ$. 
In Fig.~\ref{fig:ew_with_cartoon} we show the actual predicted histograms 
from the toy model for four different values of $\theta_{\mathrm{b}}$ and $\Delta \theta = 20^\circ$, 
with an accompanying cartoon showing the viewing angles to BAL quasars in each case. Noticeable
differences in both the peaks and widths of the distributions are present for equatorial models.
Together, Figs.~\ref{fig:deltamu} and \ref{fig:ew_with_cartoon} show that, 
under the assumptions of our toy model, geometries in which BAL quasars 
are viewed from similar (i.e. low) inclination angles 
to non-BAL quasars are strongly favoured. 

In the context of geometric models, the simulations can be reconciled with
observations via a number of possible scenarios linked to the assumptions made, namely:
\begin{itemize}
\item BALQs are viewed from angles comparable to non-BAL quasars, i.e. low inclinations.
\item For reasons that are unclear, quasar disc emission is roughly isotropic, i.e. $\epsilon(\theta)\sim 1$.
\item The line emission is strongly anisotropic in the same fashion as the continuum, i.e. 
$\epsilon_{\mathrm{line}}(\theta) \sim \epsilon(\theta)$. 
\item Other factors, such as obscuration or selection effects in the sample, 
are hiding the expected behaviour.
\end{itemize}
Clearly, a combination of two or more of these effects is also plausible.
These scenarios are explored further in the remaining sections.

 %fullpage

\subsection{The Effect of Flux Limits}
\label{sec:flux_select}

In the above procedure, inclinations are generated isotropically,
so that the probability of a given observer orientation is simply proportional
to solid angle. In reality, an anisotropic continuum source
will cause substantial bias towards low inclination sources in
flux-limited samples. The effect of flux limits
on the expected distributions of angles and EWs, 
as well as the BAL fraction, $f_{\mathrm{BAL}}$, 
is interesting. Imposing such a 
limit in the above analysis means fewer high inclination
objects with high EWs will appear in the mock sample. The value of
$\Delta \mu_{\mathrm{EW}}$ will therefore decrease for a 
given bending angle. Furthermore, as noted by \cite{krolik1998},
the covering factor of the outflow must dramatically 
increase as $\theta_b$ increases to reproduce the observed value
of $f_{\mathrm{BAL}}$.

We have conducted some preliminary tests, which show 
that the distribution of observer orientations in a mock sample is
very sensitive to the flux limit used;
the choice of flux limit to impose is thus both crucial 
and non-trivial. Ideally, one would reconstruct the 
quasar luminosity function for each geometry by deconvolving 
the {\em observed} flux distribution from the model angular 
distribution. Samples could then be drawn for each 
generated angle,
corrected by $1/\epsilon(\theta)$, and then required to pass the 
magnitude limit of the actual sample. We reserve this process,
and an investigation of the complex effects on the true value
of $f_{\mathrm{BAL}}$, for a future study.
Fortunately, the conclusions of this paper are not 
particularly sensitive to this issue, as the sharper the imposed
flux cutoff, the more equatorial BAL outflows are prohibited by the model.
This leads to similar conclusions as those drawn from the
similarity in EW distributions: a more isotropic disc or a non-equatorial 
viewing angle for BAL quasars are favoured. 
The bending angle, $\theta_b$ can thus be thought of as the
flux-weighted average viewing angle -- this is, after all, the 
angle that is really being inferred from any orientation indicator
with similar selection biases to the SDSS.

% %%%%%%%%%%%%%%%%%%%%%%%%%%%%%%%
% % DISCUSSION
% %%%%%%%%%%%%%%%%%%%%%%%%%%%%%%%

\section{Discussion}
\label{sec:discuss}
We have demonstrated that the EW distributions of the emission lines in 
BAL and non-BAL quasars are not consistent with a 
model in which BAL quasars are viewed from equatorial angles 
and the continuum emission originates from an optically thick, geometrically thin
accretion disc. A number of simplifications were made in the models presented in 
section~\ref{sec:toymods}: the full effects of both general relativity (GR) and frequency-dependent opacities in the disc were ignored, 
emission line isotropy was assumed and there was no modelling of obscuration of the continuum 
source. We therefore now discuss the potential impact of each of these effects.
We generally focus on the distribution of \ewo, as it is the most reliably isotropic line, but 
do devote some time to also discussing the 
broad emission lines (\civline, \ciiiline\ and \mgline).

\subsection{The Angular Distribution of Emission from an Accretion Disc}
\label{sec:disc_agn}

The most widely-used theoretical model for thin accretion discs
is the so-called `$\alpha$-disc' model of \cite{shakurasunyaev1973}
There are a number of well-documented problems when fitting 
AGN SEDs with thin disc models \cite[e.g.][]{koratkar1999,antonucci2013,shankar2016}. 
Despite these problems, \cite{capellupo2015} succeeded
fitting $\alpha$-disc models to AGN spectra when the effects of
GR, mass-loss and comptonisation were included.
In this section, we start by discussing the angular distribution of
emission from a classic $\alpha$-disc, before exploring opacity and GR 
effects. In order to do so, we use \agn\
\citep{hubeny2000,davishubeny2006,davis2007}. We stress that the 
discussion here is not limited to $\alpha$-discs; the only real condition
for the angular distributions derived here is that the 
disc is geometrically thin and optically thick.

Any geometrically thin, optically thick disc will appear
foreshortened and limb darkened (if temperature decreases
with height from the central disc plane). 
Foreshortening is a simple $\cos \theta$ geometric effect. 
Limb darkening, $\eta(\theta)$, is usually approximated by a linear dependence
of the emergent flux on $\cos \theta$, i.e. 
\begin{equation}
\eta(\theta) = a \left( 1 + b \cos \theta \right),
\end{equation}
where $a$ is a normalisation constant, and $b$ governs the strength
of the limb darkening. Setting $b=3/2$
tends to give good agreement with solar observations 
\citep[e.g.][]{mihalas}. 
In reality, limb darkening is not frequency independent and 
depends on the bound-free and bound-bound opacities in the disc.
In addition, it has been shown that GR light bending can `isotropize' the radiation
field in XRBs \citep{zhang1997,munozdarias2013}, in some cases overcoming
foreshortening effects.

\begin{figure}
\centering
\includegraphics[width=0.45\textwidth]{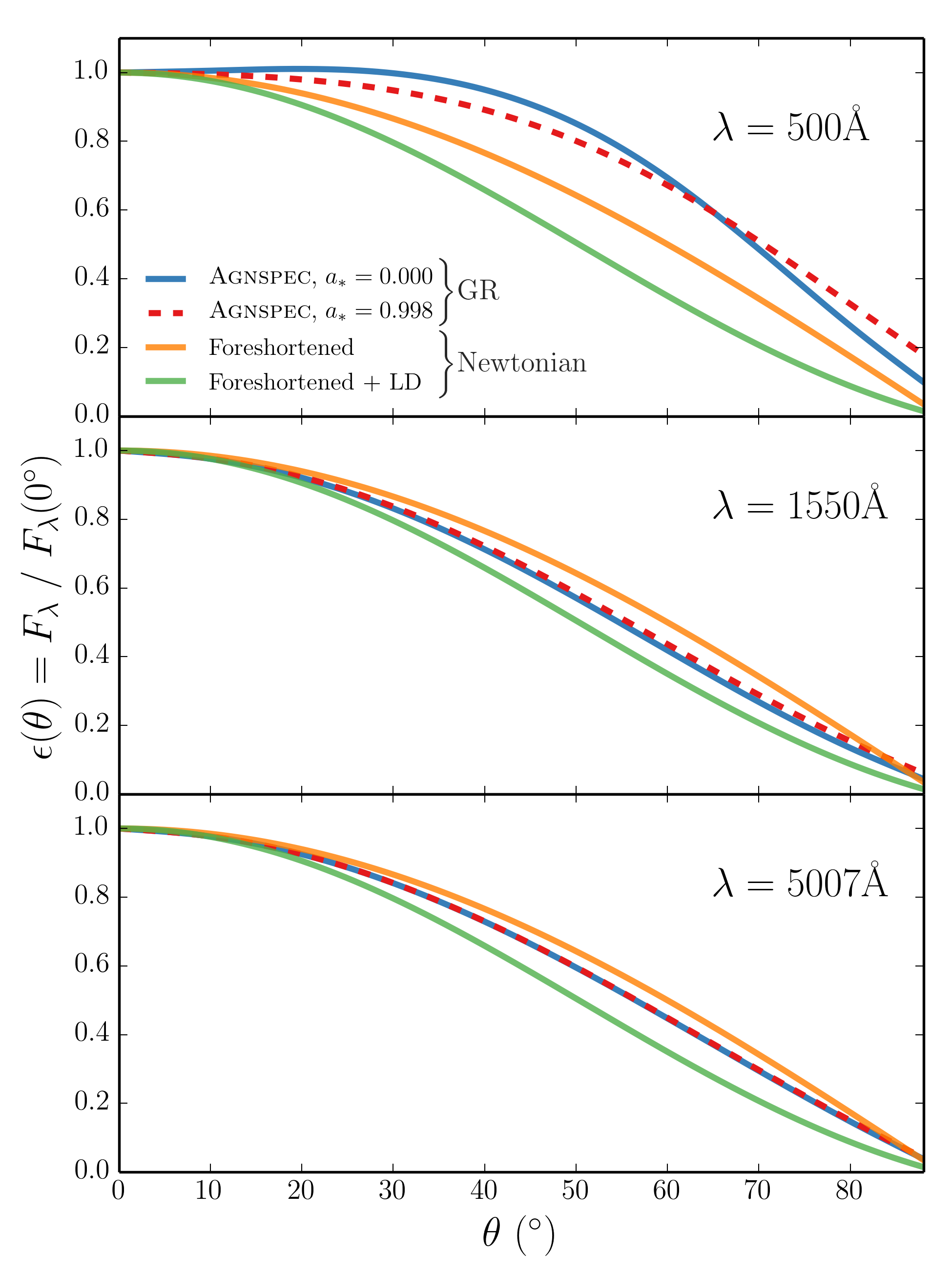}
\caption
[Angular variation of continuum luminosity from \agn\ and classical thin disc models.]
{
Angular variation of continuum luminosity from \agn\ and classical thin disc models.
The monochromatic continuum luminosities is divided by the monochromatic continuum luminosity
at $10^\circ$, from \agn\ and classical thin disc models, at three different wavelengths.
The models are computed for an Eddington fraction of $0.2$ and $M_{BH}=10^9~M_\odot$. 
In each panel we show both Kerr and Schwarzschild \agn\ models, and the classical models are
for both pure foreshortened discs and foreshortened and limb darkened (LD) discs.
}
\label{fig:agnspec_disc}
\end{figure} 

In order to assess the impact of GR and disc opacities
on \ept, we use \agn\ models. \agn\ works by first conducting a stellar atmosphere calculation
to obtain the SED from a series of annuli, before using \kerrtrans\ \citep{agol1997}
to calculate the emergent SED by ray-tracing along Kerr geodesics.
Fig.~\ref{fig:agnspec_disc} shows \ept\ as a function of 
$\theta$ for two \agn\ models for minimally and maximally spinning BHs. 
The models are characterised by $M_{BH}=10^9~M_\odot$ and an Eddington fraction of $0.2$.
The angular distribution is fairly insensitive to these choices.
For comparison, we also show foreshortened and limb-darkened predictions for SS73 models.
Although the \agn\ continua are significantly more isotropic at $500$~\AA,
there is very little effect redward of around $1000$~\AA, which is the relevant
region of \ept\ for \oiiifull, \civline, \ciiiline\ and \mgline . 
In fact, using the foreshortened estimate is the conservative (least anisotropic) prescription 
in these regimes. This therefore justifies the form of \ept\ used in the toy models
and demonstrates that GR does not affect the shapes of the emission line 
EW distributions in the UV and optical regimes. 

\subsection{Obscuration}
\label{sec:obscure}

Differential obscuration of the continuum source and 
line emitting region by a dusty torus or other circumnuclear 
absorber can change the observed EW. 
\citet[][hereafter C11]{caccianiga2011} showed that the distribution of \ewo\
can also be well fitted by an obscuration model. 
They find that AGN with column densities of 
$N_H\gtrsim10^{22}$~cm$^{-2}$ can explain the high EW powerlaw tail. 
BALQs exhibit strong X-ray absorption with column densities
of $N_H\sim10^{22-24}$~cm$^{-2}$ \citep{green1996,gallagher1999,mathur2000,green2001,grupemathur2003,morabito2013}. 
This places BALQs firmly in the EW tail according to the C11 model. 
Of course, only LoBAL quasars had \ewo\ measurements 
in the sample used here, but these generally
show even higher column densities, approaching Compton-thick values 
\citep{morabito2011}.

We therefore suggest that the obscuration model of C11 cannot explain the \ewo\ distribution
of LoBALQs. The similarity of the observed LoBAL and non-BAL distributions
also means that obscuration is unlikely to drive the behaviour of \ewo\ as a whole.
These conclusions are moderated if the line of sight to the X-ray source, which determines 
the measured $N_H$, experiences a different absorbing column to that of the optical continuum.
They are also dependent on the particular absorption model used by C11. Indeed, it is worth noting at 
this point that there is a degree of scatter in the $N_H$ values measured from X-ray and optical
observations \citep{maiolino2001b,maiolino2001a}, as could be produced by differing 
viewing angles to the X-ray and optical radiation sources. Further work is clearly needed, but considering the absorption properties of BALQs strengthens the findings presented in section~\ref{sec:results}.

\subsection{Line Anisotropy}
\label{sec:line_aniso}

Optically thin lines are isotropic -- the {\em local}
escape probabilities in each direction are equal due to the 
low optical depth. Anisotropy can however be introduced into optically thin 
line emission by variation in continuum absorption. Indeed,
\cite{kraemer2011} showed that the strength of \oiiifull\ compared
to the infra-red [O\textsc{iv}]~$28.59~\mu$m line 
varies between type 1 and type 2 AGN. However, this variation is due 
to frequency-dependent absorption, so should not affect the distribution of \ewo.
If a higher continuum optical depth was experienced along the line of sight
to the NLR than to the continuum source then this could mask EW
trends, but this is the opposite behaviour than that expected from type 1/type 2
unification geometries.

When lines are optically thick, the situation is more
complex, as local velocity gradients then determine their
anisotropy. 
Keplerian velocity shear has been shown to modify the
shape of disc-formed emission lines \citep{hornemarsh1986}, 
whilst an additional
radial shear from a wind can cause double-peaked lines
to become single-peaked \citep{MC96,MC97,flohic2012}. 
Although there is a sub-population of AGN with double-peaked lines 
\citep[e.g.][]{eracleous1994,eracleous2003},
this fraction is only around $3\%$ \citep{strateva2003}, so
AGN and quasar spectra in general show broad, single-peaked lines.
The single-peaked nature of most quasar emission lines either implies 
that they are not formed in a Keplerian disc, that quasars are mostly 
viewed pole-on, that radial velocity gradients modify the profile shapes, 
or that an additional single-peaked component is required \citep{storchi2016}. 
Disc-shaped BLRs are popular in the literature 
\citep[e.g.][]{decarli2008,gaskell2013,begelmansilk2016}
and are commonly invoked to explain reverberation mapping results 
\citep{nunez2013,pancoast2014a,pancoast2014b,goad2014}.
Furthermore, R11 suggested that the broad emission lines 
form in a disc and 
trace the disc emission in terms of their anisotropy. 
If this was the case, we would not expect a difference in 
the BAL and non-BAL quasar EW distributions. 
However, an emission line is only purely 
foreshortened if formed in a disc with zero velocity shear or an 
isotropic local velocity gradient. Neither of these scenarios
are particularly plausible.

We have explored the expected angular distributions
of line emissivity if the lines came from a region subject to 
Keplerian velocity shear. 
A more detailed discussion is found in Appendix A. 
The results show that optically thick line emission 
from a Keplerian disc 
does not follow a $\cos \theta$ unless $H/R\sim0.01$,
which is unrealistically small for the 
BLR \citep{kollatschny2013,pancoast2014b}
Thus, a {\em Keplerian} disc-like BLR cannot explain 
the overall EW distributions 
of the broad emission lines or the similarity 
of the distributions of \civline\ EW and \mgline\ EW in BAL and 
non-BAL quasars. The presence of an equatorial wind would 
exacerbate the effect and cause more 
line emission to escape along the radial velocity 
gradient towards high inclinations.
Other models, in which the BLR is made up of a 
series of clouds orbiting in a disc-shaped structure 
\citep[see][ for a review]{sulentic2000bels}
could feasibly produce a $\cos \theta$ 
angular dependence.
In this case local line emission
is isotropic but is then affected by continuum opacity 
from other orbiting clouds. Radiative transfer effects 
within the cloud could also lead to 
more complex line anisotropy effects 
\citep{davidson1979,ferland1992,korista1997}.
We cannot exclude such a model but argue that, as line 
anisotropy cannot significantly affect the distribution 
of \ewo, it is also unlikely to significantly affect 
even the EW of lines formed in the BLR from permitted 
dipole transitions.

\begin{figure*}
\centering
\includegraphics[width=0.75\textwidth]{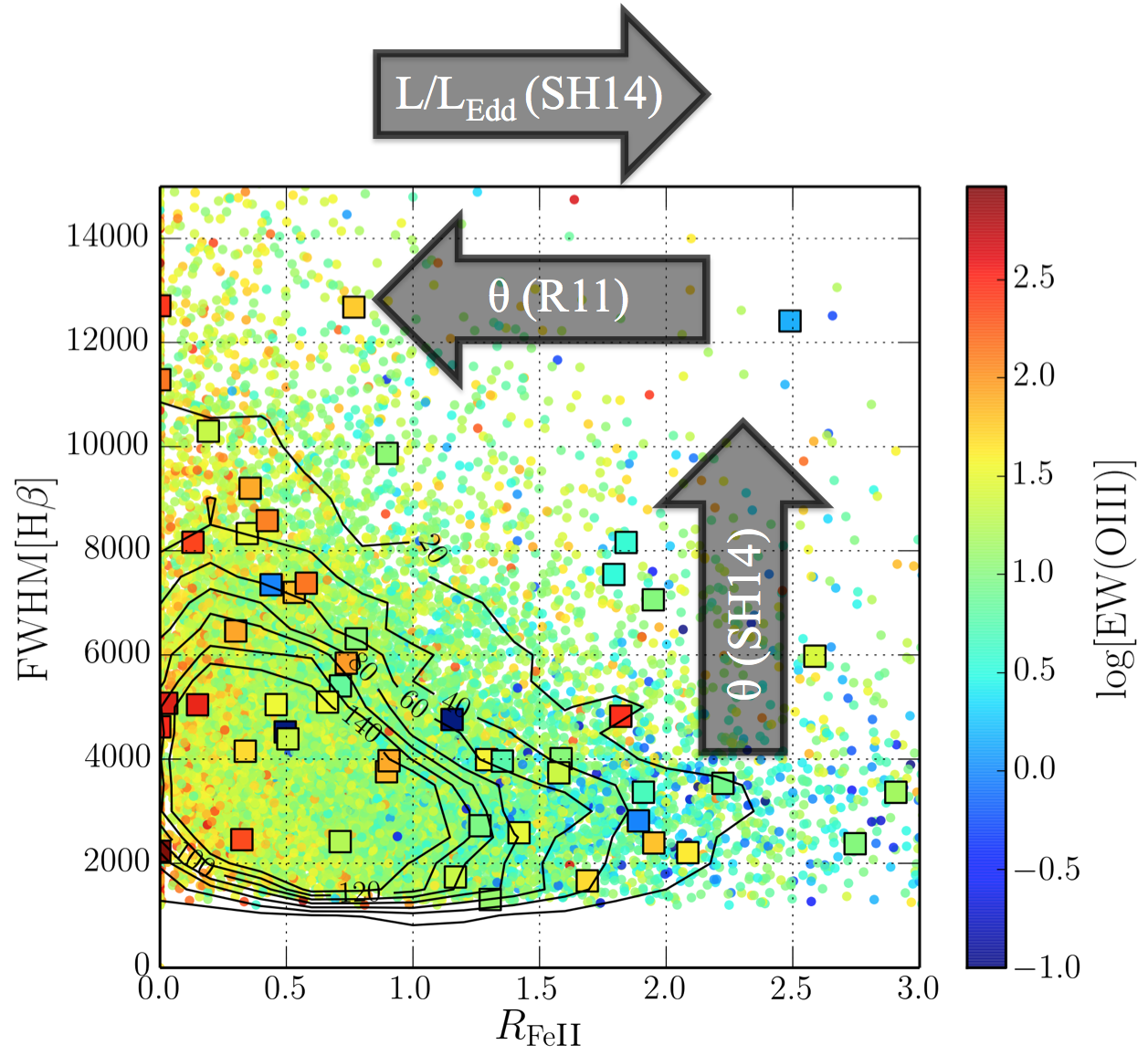}
\caption
[Eigenvector 1 for LoBAL and non-BAL quasars.]
{
Eigenvector 1 for LoBAL and non-BAL quasars. 
FWHM of the \hb\ line plotted against the relative
iron strength, $R_{{\rm Fe \textsc{ii}}}$. The colour coding
corresponds to \ewo. The dots mark all quasars from
sample A, while the squares mark those with \mgii\ LoBALs. 
A few of the \mgii\ LoBALQs are missing due to their lack of \fwh\ 
measurements. The arrows show the approximate 
direction of the expected trend with increasing inclination ($\theta$)
under both the SH14 and R11 interpretations, and the expected 
trend with increasing Eddington fraction ($L/L_{\mathrm{Edd}}$) from SH14 only.
HiBAL quasars cannot be placed on this plot due to the lack of rest-frame
optical coverage.
}
\label{fig:bal_ev1}
\end{figure*}

\section{Clues from Other Sources}

Having explored the behaviour of emission line EW 
as a potential orientation indicator, we now devote some
time to discussing other observables that might trace viewing angles
to BAL and non-BAL quasars. We will also briefly discuss the constraints 
from theoretical work and the potential implications for the wind-driving mechanism.

\subsection{Eigenvector 1}

Eigenvector 1 (EV1) is a fundamental parameter space for AGN and quasars
\citep{borosongreen,sulentic2000ev1,marziani2001,shenho2014}. 
It relates the FWHM of \hb, the relative iron strength, 
$R_{{\rm Fe \textsc{ii}}}$, and
\ewo. Both \ewo\ and \fwh\ have been used as orientation
indicators: \fwh\ should increase with inclination if the line formation
region is at all disc-shaped due to velocity projection effects.
This means that comparing the LoBALQ EV1 distribution to the non-BAL 
quasar EV1 distribution is particularly interesting. Once again,
the SDSS HiBALQs cannot be placed in this space due to the lack of rest-frame 
optical coverage.

Fig.~\ref{fig:bal_ev1} shows the quasar distribution from sample A 
in EV1 parameter space, with LoBAL quasars from sample A overplotted.
\citet[][hereafter SH14]{shenho2014} propose 
that the main inclination driver in this parameter space
is \fwh, and that high inclination sources should thus cluster nearer 
to the top of the plot. In contrast,
R11's analysis predicts that high inclination sources should cluster
around high \ewo. As \ewo\ and \fwh\ are very weakly correlated
(Spearman's rank coefficient of 0.14), this means they should lie to
the left of the parameter space, due to the clear correlation between \ewo\
and $R_{{\rm Fe \textsc{ii}}}$. These expected trends are shown with arrows
in Fig.~\ref{fig:bal_ev1}; inspection of the figure clearly 
shows that LoBAL quasars are not confined to one region of the 
EV1 parameter space. This is contrary to previous findings with different samples
in which LoBAL quasars were thought to lie at extreme ends due to low \ewo\ and high 
$R_{{\rm Fe \textsc{ii}}}$ \citep{boroson1992,turnshek1997,runnoe2014}. 

\begin{figure}
\centering
\includegraphics[width=0.5\textwidth]{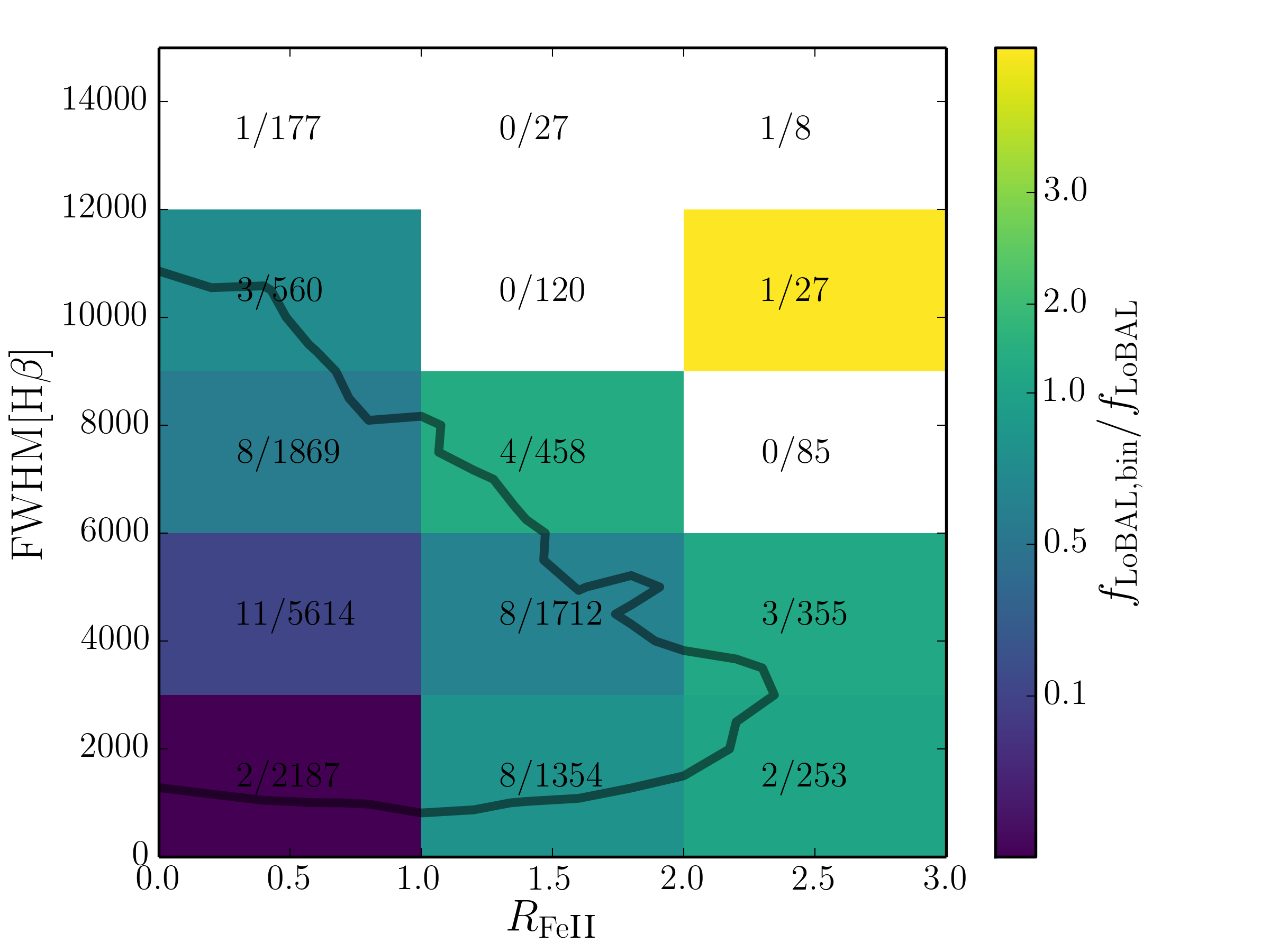}
\caption
[LoBAL fraction compared to global LoBAL fraction in Eigenvector 1 space.]
{
LoBAL fraction compared to global LoBAL fraction in Eigenvector 1 space, in bins
of $\Delta R_{{\rm Fe \textsc{ii}}} = 1$ and $\Delta$\fwh$=3000$km~s$^{-1}$..
The contour shows the outermost contour from Fig.~\ref{fig:bal_ev1} for
reference. The text shows $N_{\mathrm{LoBAL}}/N_{\mathrm{non-BAL}}$, 
where $N_{\mathrm{LoBAL}}$ is the number of LoBALQs in the bin and 
$N_{\mathrm{non-BAL}}$ in the number of non-BAL quasars in the bin.
}
\label{fig:bal_ev1_bins}
\end{figure}

In order to assess this more quantitatively, we also show contours of 
quasar counts overlaid on the scatter plot. The contours correspond
to the number of objects in each bin, where the bins are of size
$\Delta R_{{\rm Fe \textsc{ii}}} = 0.2$ and $\Delta$\fwh$=500$km~s$^{-1}$.
The percentage of quasars falling within the inner contour is 45\%, 
whereas only 18\% of LoBALQs fall in the space. Conversely, 24\% 
of LoBALQs fall outside the outermost contour compared to 10\% of 
non-BAL quasars. It would therefore appear that BAL 
quasars are slightly preferentially clustered towards the high-mass and 
high-inclination end of EV1 space (under the interpretation of SH14).
This is further illustrated by Fig.~\ref{fig:bal_ev1_bins},
which shows the LoBAL fraction in larger bins, compared to the 
mean LoBAL fraction. This is again suggestive of an overdensity of LoBALQs 
towards the upper right of the parameter space.
It is also clear that a unification picture in which BAL 
quasars are viewed exclusively from high inclinations is 
inconsistent both the R11 and SH14 interpretations of EV1 parameter 
space. 

Larger datasets, preferably including HiBAL quasars with EV1 measurements, 
are needed in order to properly constrain the EV1 behaviour of BAL quasars.
However, overall, the behaviour of EV1 in LoBALQs slightly 
strengthens the conclusion that BALQs are not viewed exclusively from 
extreme inclinations, or alternatively, that we do
not yet understand the real drivers of EV1.

\subsection{Polarisation}

\begin{figure*}
\centering
\includegraphics[width=0.99\textwidth]{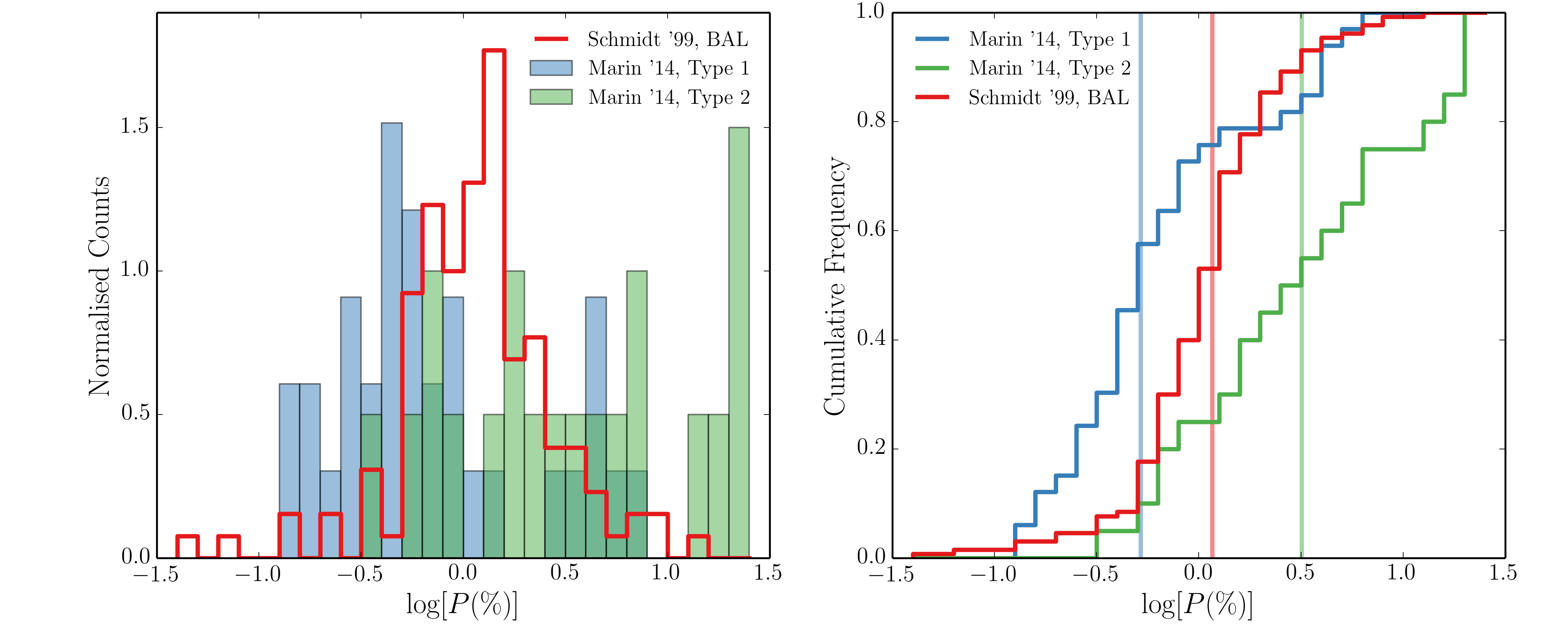}
\caption
{
Left: Histograms of polarisation percentages 
for BAL quasars from Schmidt et al. (1999) together with the 
Marin et al. (2014) AGN sample. 
Right: Cumulative distribution functions of the histograms shown in
the left panel, with the same colour-coding and $x$-axis scale. 
The translucent vertical lines mark the median value in each sample.
}
\label{fig:bal_polarisation}
\end{figure*}

Spectropolarimetry of BAL quasars offers some of
the best insights into the geometries of BAL outflows and 
tends to show a few key properties. The first is enhanced 
polarisation in the BAL troughs themselves \citep{schmidt1999,ogle1999}. 
This is readily explained by a scattering region unobscured by the
BAL trough, with the higher polarisation percentage simply due to the
decreased direct flux. The second property is a continuum 
polarisation percentage in BALQs that is around $2$ times greater, 
on average, than seen in the non-BAL population \citep{schmidt1999}.
A histogram of the continuum polarisation percentages 
of a sample of BAL quasars from 
\cite{schmidt1999} are compared to the type 1 and type 2 AGN 
populations from \cite{marin2014} in the left panel of Fig.~\ref{fig:bal_polarisation}.
The corresponding cumulative distribution function is shown in the 
right panel. These show that BAL polarisation percentages
lie between those of type 1 and type 2 AGN. 
If type 1 and type 2 objects are viewed from low and high 
inclinations, respectively, as expected from unified models, 
this implies intermediate inclinations for BALQs.
%from unified models.

The third characteristic polarisation property of BALQs 
is a polarisation angle with respect to the radio axis, 
$\Delta \mathrm{PA}$ of $\gtrsim60^\circ$ in RL objects, 
as found in all of the seven BALQ
measurements compiled by \citet{brotherton2006}.
These observations have been explained
by a model with a polar scattering region, distinct from the BLR and 
BAL regions, which is then viewed at an equatorial angle 
\citep[e.g.][]{goodrich1995, cohen1995,lamy2004}. 
However, large values of $\Delta \mathrm{PA}$
can be produced at intermediate viewing angles
of around $45^\circ$ \citep{kartje1995,smith2004,axon2008,borguet2011}.
We suggest that polarisation predictions similar to those described by
\cite{marin2013} are carried out for full radiative transfer wind 
models \citep[e.g.][]{M16}. Overall, however, quasar 
polarisation properties imply that 
BALQs are viewed from higher-than-average, intermediate viewing 
angles between type 1 and 2 AGN.

\subsection{Radio Properties}

Radio measurements have long been proposed as a potential probe of the 
quasar orientation \citep{orr1982}. In particular, measurements of jet Lorentz 
factors, radio spectral index and the ratio of the core flux to extended flux, the so-called `radio core dominance', 
$\log R$, offer ways to measure radio jet inclination. 
Although they are preferentially 
radio-quiet \citep{stocke1992,brotherton1998,shankar2008},
radio observations of BALQs 
offer some of the strongest evidence against equatorial geometric models: for one, they have 
lead to the discovery of the (aforementioned) polar BAL quasars 
\citep{zhou2006,ghoshpunsly2007,berrington2013}. While there is evidence that 
BALQs generally possess steeper spectral indices \citep{dipompeo2011_vla,bruni2012}, 
the range of values is indicative of a variety of viewing angles
\citep{becker2000,montenegro2008,dipompeo2012b}. 

Recent long baseline radio observations of BALQs 
% have also been made 
% \citep[e.g.][]{jiang2003,kunert2007,kunert2010}. 
%The more recent of these 
found very compact ($<10$~pc) radio structure, 
suggestive of a young radio source \citep{doi2013,kunert2015}, 
although orientation is also thought to have a role 
\citep{ceglowski2015}. 
The similarity in UV properties means
that many of the conclusions drawn from radio-loud samples should also 
be applicable to radio quiet BALQs \citep{rochais2014}.
Overall, radio studies of BALQs imply that they are viewed from a 
range of angles but higher than average, 
as suggested by, e.g., \cite{dipompeo2011}. 
Neither evolutionary or geometric models
can adequately account for the global radio properties of 
BALQs \citep[see ][for a good discussion]{doi2013}.

\subsection{Theory of Winds}
\label{sec:wind_theory}
Theoretical predictions for outflow geometries depend on the 
driving mechanism being considered. 
Two of the most promising ways for BAL outflows to be launched are via 
the magnetocentrifugal `bead on a wire' mechanism \citet{blandfordpayne} 
or radiation pressure on spectral lines (`line-driving'). 
Given the lack of knowledge about the magnetic field 
in quasar accretion discs, theoretical 
considerations provide very few constraints in the case of magnetocentrifugal outflows. 
Tests of the \cite{emmering1992}
model, in which a magnetohydrodynamic wind
produces the broad emission lines,
favour launching angles of $\gtrsim 30^\circ$
in order to match observed linewidths
\citep{chajet2013,chajet2016}. 
However, the exact relationship between such a wind and 
BAL outflows is not clear.

We know line-driving is present in many quasar outflows thanks 
to the discovery of line-locking signatures 
(\citealt{arav1995,arav1996,north2006,bowler2014}; 
but see also \citealt{cottis2010}).
Hydrodynamic simulations have been successful in producing 
substantial line-driven disc winds \citep{PSK2000,PK04}.
Although these models generally favour equatorial flows, 
the ionization state is crucial in determining the degree of 
collimation and depends on disc and X-ray luminosities as well as 
black hole mass \citep{proga1998,PK04}. 
The level of shielding \citep{MCGV95,PK04}
and clumping \citep[e.g.][]{junk1983,weymann1985,hamann2013,M16}
may also be important. Fully coupled radiative transfer and 
hydrodynamic simulations are important in this regard as they
properly account for the radiation field \citep{simproga2012,H14}.

% %%%%%%%%%%%%%%%%%%%%%%%%%%%%%%%
% % CONCLUSIONS
% %%%%%%%%%%%%%%%%%%%%%%%%%%%%%%%

\section{Conclusions}
\label{sec:ew_conclusions}

We have explored the emission line properties of BAL and non-BAL quasars,
particularly focusing on the EW distributions in two redshift ranges of the SDSS
quasar catalog. Our main conclusion is that the EW distributions of BAL and
non-BAL quasars are remarkably similar and that this is {\em not} what one would expect from
a unification model in which an equatorial BAL outflow rises from a foreshortened 
accretion disc. This geometry is widely used in 
geometric unification and BAL outflow models
\citep[e.g.][]{MCGV95, dekool1995,elvis2000, PSK2000,risalitielvis2010, PK04, borguet2010,
higginbottom2013, nomura2013,nomura2016}.

In order to calculate the expected distributions from different wind geometries,
we conducted a series of simulations similar to those described by \cite{risaliti2011}.
As expected, these simulations confirmed the above finding, predicting
differences of $>10$\AA\  in the mean EW for $\theta_{\mathrm{b}} \gtrsim 60^\circ$ -- differences which are not seen in the data. 
We demonstrated that GR or opacity effects in the disc do not cause the 
continuum to become more isotropic in the relevant wavelength regimes. 
Line anisotropy and obscuration cannot effectively hide
the expected inclination trend -- in fact, the column density measurements
from X-ray observations of BAL quasars suggest that 
obscuration does not drive the distribution of \ewo.

There are three basic ways to explain our results:
\begin{itemize}
	\item {\sl Scenario A:} 
    Quasar discs are radiate anisotropically, as expected from a 
	geometrically thin, optically thick accretion disc. In this case, BAL outflows cannot 
	only emerge at extreme inclinations and should instead be seen from low to 	intermediate inclinations.
    \smallskip
	\item {\sl Scenario B:} The quasar continuum is much more isotropic than one would
	expect from a geometrically thin, optically thick accretion disc.
	\smallskip
	\item  {\sl Scenario C:} The geometric unification model does not explain the incidence of 
	BALs in quasars, or requires an additional component which is {\em time-dependent}, 
	such as an evolutionary or accretion state origin for BAL outflows. If this is the case,
	the wind must still have a geometry, which is important to constrain in order to understand
	the driving mechanism and estimate the feedback efficiency.
\end{itemize}

We then examined the relative merits of these scenarios 
in the context of the large body of work on quasar 
orientation and BALQs. Based on the polarisation and 
radio properties, a scenario in which 
BAL outflows emerge at intermediate inclinations between 
type 1 and type 2 AGN is favoured. It is still 
possible that the actual outflow has a wide covering 
factor, but lower inclination quasars appear
preferentially in flux-limited samples due to the effect 
of foreshortening and continuum absorption. We recommend 
that future radiative transfer modelling efforts 
explore different outflow geometries 
and that detailed polarisation modelling is undertaken 
to constrain the outflow opening angles. We note that
the current benchmark line-driven wind models produce 
equatorial flows (see section~\ref{sec:wind_theory})
-- a picture that is in tension with our results.

We also suggest that the accretion disc physics is a 
crucial aspect to understand, 
particularly if equatorial winds are indeed present. 
We cannot distinguish more effectively between the above 
scenarios until the continuum 
source in AGN is understood. We point to the so-called 
`accretion-disc size problem' 
\citep[e.g.][]{morgan2010,edelson2015}
as one of many results that indicates further problems 
when reconciling thin disc 
quasar models with observations. We hope that a "two 
birds, one stone" scenario 
may arise that explains some of the observed 
discrepancies simultaneously. 
If nothing else, our work adds to the evidence that many 
simple unification models are not sufficient to describe 
the diverse phenomenology of AGN and quasars.

\section*{Acknowledgements}

The work of JHM is funded by the Science and Technology Facilities Council (STFC),
via a studentship whilst at Southampton and a consolidated grant at Oxford. 
The work of CK is also supported by the
STFC and a Leverhulme fellowship. KSL acknowledges the support of NASA for this work through grant NNG15PP48P to serve as a science adviser to the Astro-H project.
We would like to thank the anonymous referee for a constructive and insightful report that improved the quality of the paper.
JHM would particularly like to thank Lance Miller and 
Poshak Gandhi for being 
such wonderful PhD examiners, and 
everyone at University of Southampton astronomy 
department wherethe bulk of this work was carried out. 
More generally, we would like to thank 
Omer Blaes, Ivan Hubeny 
and Shane Davis for their help with \agn, and 
Mike Brotherton, Mike DiPompeo and Frederic Marin for 
their correspondence regarding
polarisation measurements and orientation indicators.
We are also grateful to Stuart Sim, Nick Higginbotton, 
Sam Mangham, Francesco Shankar, Daniel Proga, Leah Morabito, 
Rob Fender, Tony Bell, Katherine Blundell, Will Potter,
Sam Connolly and Sebastien Hoenig
for useful discussions. 
Figures were produced using the {\tt matplotlib} 
plotting library \citep{matplotlib}. 
This work made use of the Sloan Digital Sky Survey.
Funding for the Sloan Digital Sky Survey has been provided by
the Alfred P. Sloan Foundation, the U.S. Department of Energy Office of
Science, and the Participating Institutions.

\bibliography{paper}

\newpage
\clearpage
\appendix
\section{The Angular Emissivity Distribution
from a disc-shaped line formation region subject to Keplerian velocity shear}

In the case of an optically thick line formed in a disc-shaped line formation region subject to Keplerian velocity shear, the surface brightness is \citep{hornemarsh1986}
\begin{equation}
J_{\mathrm{thick}}(\theta) \approx \cos \theta~S_L~\Delta \nu~\sqrt{8 \ln \tau_0} \ ,
\end{equation}
where $S_L$ is the line source function (assumed constant) and
$\tau_0$ is the line centre optical depth, given by
\begin{equation}
\tau_0 = \frac{{\cal W}}{\sqrt{2}\pi \Delta \nu \cos \theta}.
\end{equation}
The parameter ${{\cal W}}$ is given by
\begin{equation}
{{\cal W}} = \frac{\pi e^2}{m_e c}f N^\prime,
\end{equation}
where $f$ is the oscillator strength and $N^\prime$ is the number
density integrated along the vertical height of the disc shaped emitting region. The linewidth
$\Delta \nu$ is enhanced from the thermal line width by the velocity shear, such
that
\begin{equation}
\Delta \nu = \Delta \nu_{th} \left[1 + 
\left(\frac{3}{4}\frac{v_{k}}{v_{th}}\frac{H}{R}\right)^2
Q(\theta, \phi)
\right]^{1/2},
\end{equation}
where we have defined
\begin{equation}
Q(\theta, \phi) =
\sin^2 \theta \tan^2 \theta \sin^2 2 \phi.
\end{equation}
Here, $\phi$ is the azimuthal angle in the disc, $\nu_{th}$ and $v_{th}$ are the 
thermal line widths in frequency and velocity units respectively, 
$H$ is the scale height of the disc at radius $R$, and $v_k$ is the
Keplerian velocity. The outcome of the \cite{hornemarsh1986} analysis is that optically thick lines 
formed in a Keplerian disc are strongly anisotropic, but they do not follow a simple 
$\cos \theta$ distribution. Instead, the line anisotropy is
a function of the velocity shear in the disc, the atomic physics of
the line in question, the location of the line formation region 
and the vertical disc structure. 

\begin{figure}
\centering
\includegraphics[width=0.48\textwidth]{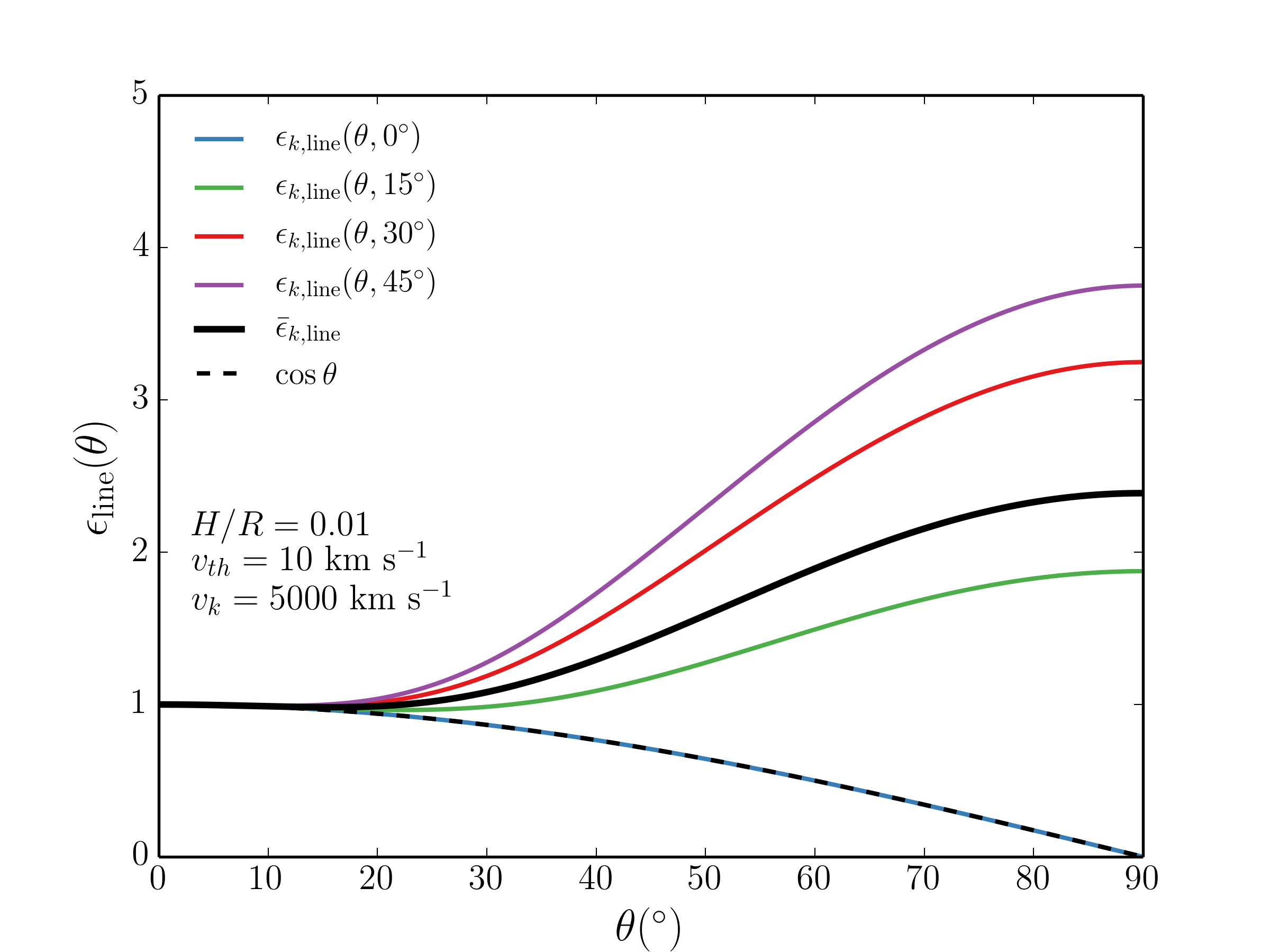}
\caption
[The line angular emissivity function from a Keplerian disc 
as a function of inclination angle.]
{
The line angular emissivity function, $\epsilon_{k,\mathrm{line}}(\theta,\phi)$,
from a Keplerian disc 
as a function of inclination angle, $\theta$, for a few different azimuthal angles, $\phi$.
The azimuthally-averaged case, $\bar{\epsilon}_{k,\mathrm{line}}$ (thick black line),
and the zero Keplerian velocity shear case, 
$\epsilon_{0,\mathrm{line}}(\theta) = \cos \theta$ (dotted line), are also shown.
}
\label{fig:line_aniso1}
\end{figure}

\begin{figure}
\centering
\includegraphics[width=0.48\textwidth]{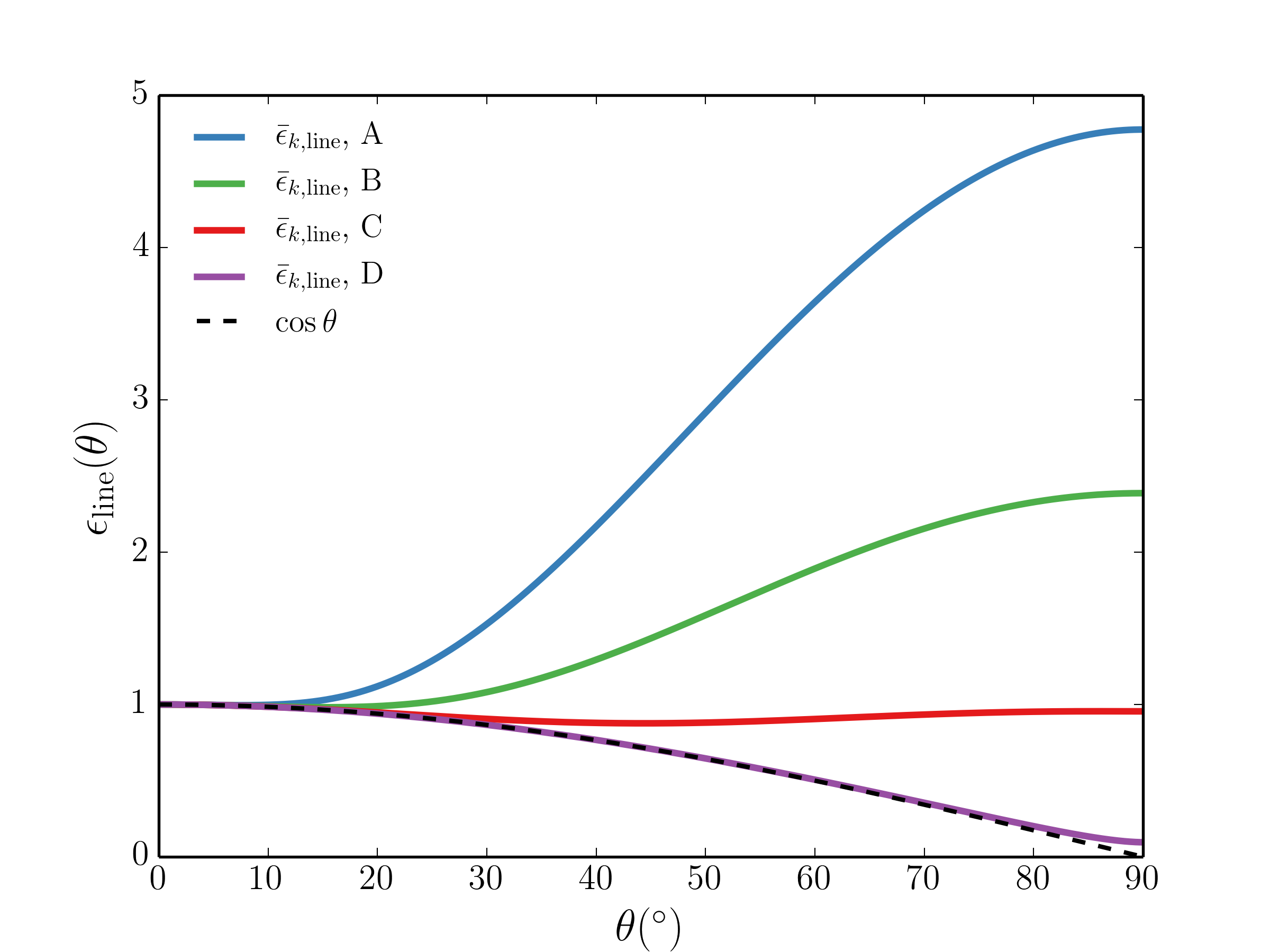}
\caption
[The line angular emissivity function from a Keplerian disc 
as a function of inclination angle.]
{
The azimuthally-averaged line angular emissivity function, $\bar{\epsilon}_{k,\mathrm{line}}$,
from a Keplerian disc as a function of inclination angle for the four models shown in 
table~\ref{tab:line_mods}. The model parameters are the values of $H/R$, $v_k$ and $v_{th}$. 
The zero Keplerian velocity shear case, $\epsilon_{0,\mathrm{line}}(\theta) = \cos \theta$ (dotted line), 
is also shown. Unless the disc is very thin ($H/R\sim0.001$),
$\bar{\epsilon}_{k,\mathrm{line}}$ shows large deviations from $\cos \theta$ and is significantly more isotropic.
}
\label{fig:line_aniso2}
\end{figure}

To examine the form of this line anisotropy, 
we can now define the angular emissivity function for a line, $\epsilon_{\mathrm{line}}(\theta)$.
In the optically thick case with no additional velocity shear, 
$\epsilon_{0,\mathrm{line}}(\theta) = \cos \theta$.
In the presence of Keplerian velocity shear, and neglecting the weak $\sqrt{8 \ln \tau_0}$ term, 
we can write
\begin{equation}
\epsilon_{k, \mathrm{line}}(\theta, \phi) = \cos \theta \left[1 + 
\left(\frac{3}{4}\frac{v_{k}}{v_{th}}\frac{H}{R}\right)^2
Q(\theta, \phi)
\right]^{1/2}.
\end{equation}
This quantity is compared to $\cos \theta$ in Fig.~\ref{fig:line_aniso1} 
as a function of for a few values of $\phi$, 
using typical quasar parameters of $v_k=10,000~\mathrm{km~s^{-1}}$ and 
$v_{th}=10~\mathrm{km~s^{-1}}$, and assuming $H/R=0.01$. We also show the 
azimuthally-averaged function, $\bar{\epsilon}_{k,\mathrm{line}}$, 
which determines the integrated emergent flux as a function of $\theta$.
Fig.~\ref{fig:line_aniso2} also shows $\bar{\epsilon}_{k,\mathrm{line}}$ 
for a few different model values of $v_k, v_{th}$
and $H/R$; the models are defined in table~\ref{tab:line_mods}.
Except in the case of a very thin disc ($H/R\sim0.001$), the 
line emissivity does not trace $\cos \theta$ and instead is 
generally biased towards high inclinations. This is discussed further in 
section~\ref{sec:line_aniso}.

\begin{table}
\centering
\begin{tabular}{p{1cm}p{2cm}p{2cm}p{2cm}p{2cm}}
\hline
Model & $H/R$ & $v_k (\mathrm{km~s^{-1}})$ & $v_{th} (\mathrm{km~s^{-1}})$ \\
\hline \hline 
A & $0.01$ & $10,000$ & $10$ \\
B & $0.01$ & $5,000$ & $10$ \\
C & $0.01$ & $5,000$ & $25$ \\
D & $0.001$ & $5,000$ & $25$ \\
\hline 
\end{tabular}
\caption
[The values of the Keplerian velocity, $v_k$, thermal velocity, $v_{th}$,
and ratio of disc scale height to radius, $H/R$, for four models.]
{
The values of the Keplerian velocity, $v_k$, thermal velocity, $v_{th}$,
and ratio of disc scale height to radius, $H/R$, for four models. These values
are used as inputs to calculate $\bar{\epsilon}_{k,\mathrm{line}}$ as shown in 
Fig.~\ref{fig:line_aniso2}, and model B is also used in 
Fig.~\ref{fig:line_aniso1}.
}
\label{tab:line_mods}
\end{table}

\end{document}